\documentclass[aps,pre,preprint,superscriptaddress]{revtex4-2}
\usepackage{graphicx}
\usepackage{amsmath}
\usepackage{color}
\usepackage{subfigure}
\begin{document}

\title{Flow force calculation in Lattice Boltzmann Method}

\author{Shaurya Kaushal}
\affiliation{ Engineering Mechanics Unit, Jawaharlal Nehru Centre for Advanced Scientific Research, Jakkur, Bangalore  560064, India.}

\author{Sauro Succi} 
\affiliation{IIT@La Sapienza and Research Affiliate Physics Dep. Harvard University, Cambridge, MA 02138, USA.}

\author{Santosh Ansumali}
\affiliation{ Engineering Mechanics Unit, Jawaharlal Nehru Centre for Advanced Scientific Research, Jakkur, Bangalore  560064, India.}

\begin{abstract}
 We revisit  force evaluation methodologies on rigid solid particles suspended 
 in a viscous fluid and simulated via   lattice Boltzmann method (LBM). 
 We  point out the non-commutativity of streaming and collision operators in the
 force evaluation procedure an provide a theoretical explanation for this observation. 
 Based on  this analysis ,we propose a discrete force calculation scheme with enhanced accuracy. 
 
 The proposed scheme is essentially a lattice version of the Reynolds Transport 
 Theorem (RTT) in the context of  the lattice Boltzmann formulation. 
Besides maintaining satisfactory levels of reliability and accuracy, the method also 
handles force evaluation on complex geometries in a simple and transparent way. 

We run simulations for NACA0012 airfoil for a range of Reynolds numbers ranging from $100$ to $0.5\times10^6$ and show that the current approach significantly reduces the grid size requirement for accurate force evaluation.
\end{abstract} 

\maketitle
\section{\label{sec:level1}Introduction}
 
The lattice Boltzmann method (LBM), a discrete space-time kinetic theory, has made major leaps in solving hydrodynamic problems at low Mach numbers \cite{higuera1989, higuera1989-2, chakri2013,singh2013,thampi2013, ansumali2007,succi2001, succi2018lattice,benzi1992,aidun,chen}. 
The theoretical foundation of the algorithm  is well established and boundary condition,s at the level of discrete populations, are well developed, with a number of variants ranging from the simple and efficient bounce-back (BB) boundary condition for stationary walls to the microscopic diffusive boundary condition \cite{ansumali2002}. 
An important asset of the method by now, is its ability to deal with complex shapes and 
moving boundary problems in a simplified but efficient manner.   

For such systems involving fluid-solid interaction, along with appropriate treatment of 
boundary conditions, an accurate calculation of  force (lift and drag) or torque, on the 
solid body is often crucial.  

In order to compute these hydrodynamic forces on objects, two widely used 
methodologies are: the stress integration (SI) approach \cite{li, olga, inamuro} and 
the momentum exchange algorithm (MEA)\cite{ladd}. 

In the stress integration method (as the name suggests), one computes the force  by integrating 
the stress tensor on the surface of the body. In the momentum exchange algorithm
(MEA), the force is computed by accounting for microscopic exchange of momentum  at the wall
between fluid and solid boundary, directly in terms of the discrete probability density function. 
Though both methods are well established in general, the momentum exchange method is shown 
to be more accurate than the stress integration method \cite{mei2002} at moderate 
to high Reynolds numbers.
   
The momentum exchange method allows computation of forces and torques on solid bodies 
in a simple and an effective manner by calculating the exchange of momentum between the 
solid surface and the fluid.  
Essentially,  the momentum exchange method exploit the fact  that force exerted 
on the body should be equal to the  difference between momentum carried by populations 
going into the wall and those coming back from the wall.  

\begin{figure}[htp]
	\includegraphics[scale=0.5]{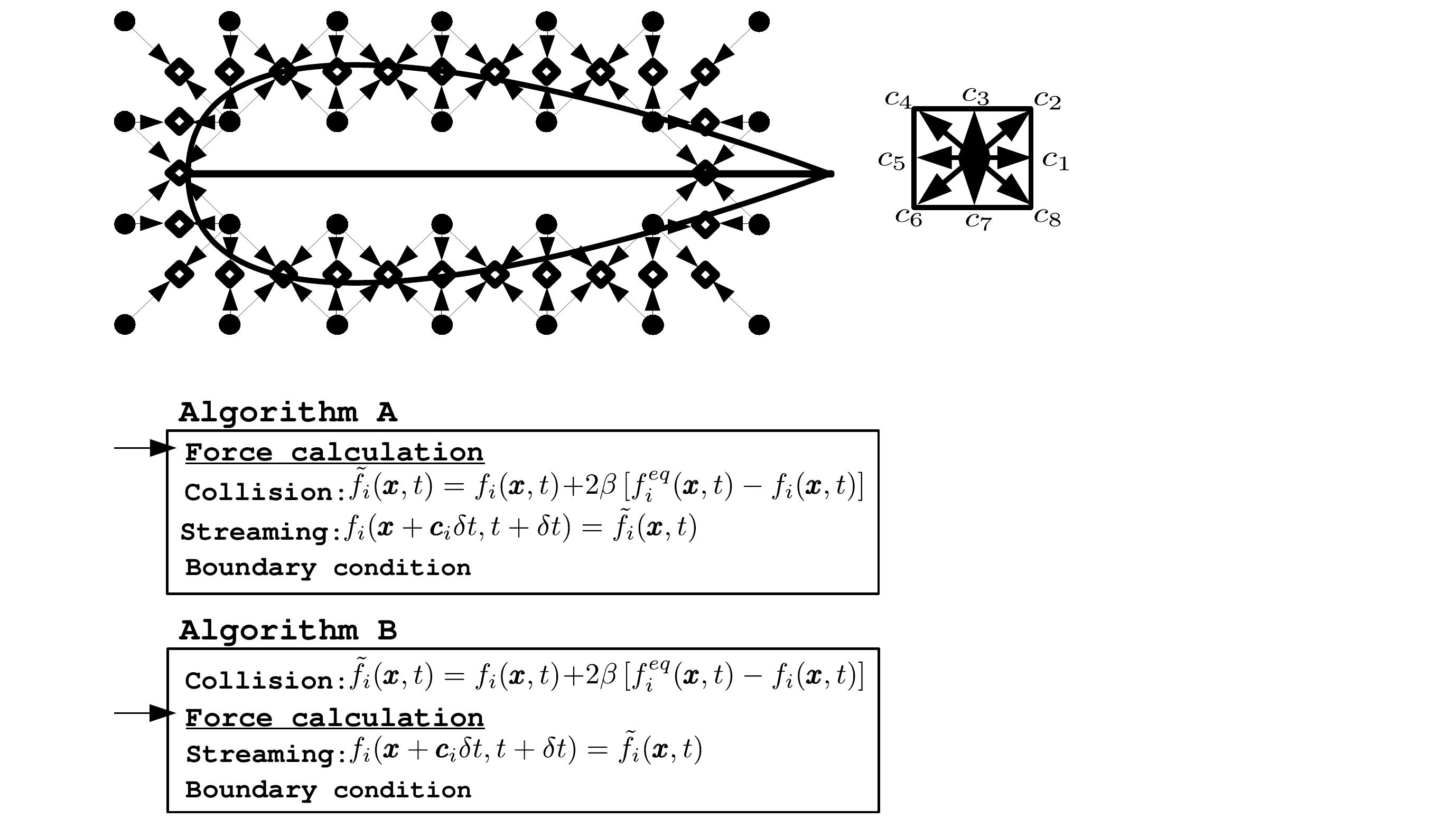}% Here is how to import EPS art
	\caption{\label{basicLadd} On top : An airfoil shaped solid object placed in a 2D cartesian grid (marked by solid circles) with a 9 velocity lattice boltzmann model. The discrete boundary points (marked by hollow squares) are placed exactly half-way between the fluid-solid links. \\
		Below : Two possible algorithms for force calculation in a lattice boltzmann code.}
\end{figure}

In its most basic version (see Fig. \ref{basicLadd}), the solid boundary is approximated in 
a discrete sense at the middle of every fluid-solid link, each of which crosses the boundary and 
connects a fluid node. Even in this case, when the boundary of solid object is laid 
down approximately, MEA is shown to be quite effective \cite{ladd, ladd1994}. 
For low resolution simulations, the disadvantage of a cartesian grid in representing a complex 
shape is often resolved either by interpolations near boundary \cite{mei2000} or by 
having multi-resolution frameworks.

\begin{figure}[htp]
	\includegraphics[scale=0.45]{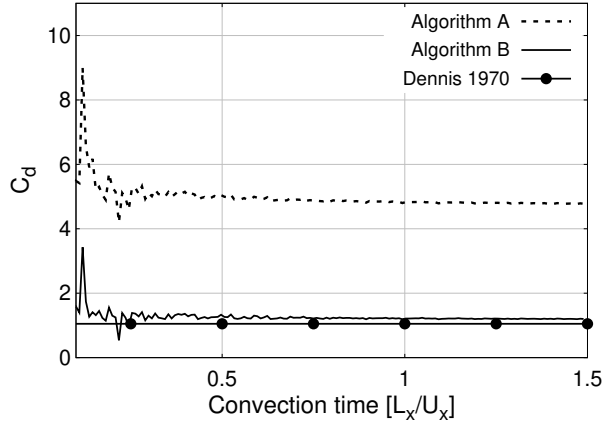}% Here is how to import EPS art
	\caption{\label{basicResults1} Coefficient of drag vs convection time for flow past a two dimensional circular cylinder of diameter $D$ solved using a D2Q9 model at Re=100 using 80 grid points per diameter. The size of the computational domain is taken to be $(50 \times 5)$D, with the cylinder symmetrically placed at $15$D from the inlet. The lattice Boltzmann results are compared with established drag values in literature \cite{dennis}.}
\end{figure}

\begin{figure}[htp]
	\includegraphics[scale=0.45]{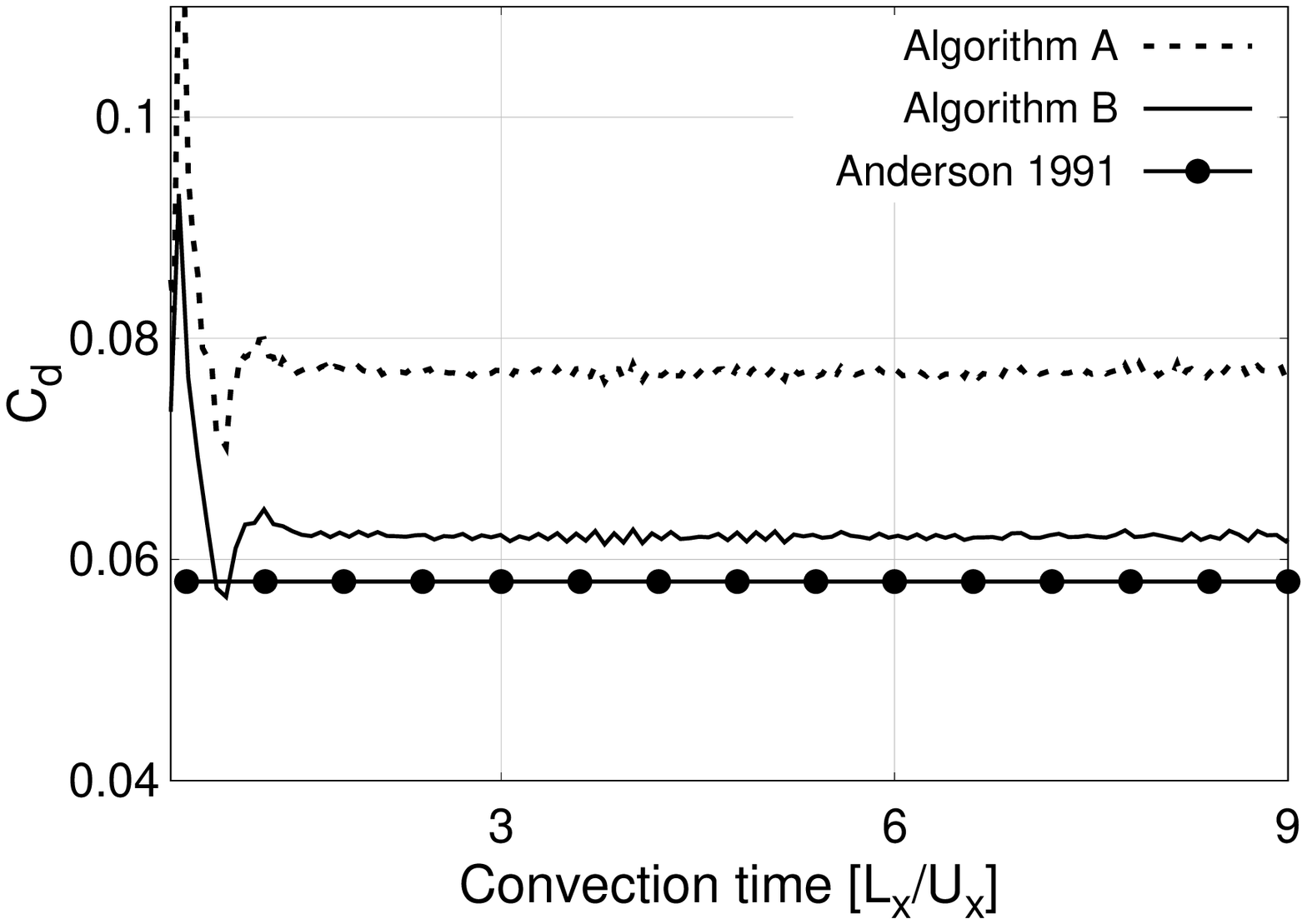}% Here is how to import EPS art
	\caption{\label{basicResults2} Coefficient of drag vs convection time for flow past a 3 dimension NACA0012 airfoil ($12\%$ thickness to chord length ratio) at AoA$=0^{\circ}$, using a RD3Q41 model at Re=5000 using 320 grid points per chord (ppc) length. The size of the computational domain is taken to be $(80\times8\times4)$ppc with the airfoil symmetrically placed at $24$ppc from the inlet.The lattice Boltzmann results are compared with established drag values in literature \cite{anderson}. }
\end{figure}

In this manuscript without loss of any generality, we adhere to the simple and widely used BB treatment of the complex boundary and   point  out a source of ambiguity in the force computation routine.  
Essentially, the issue is related to the fact that LBM is a repeated sequence of collision and streaming. In absence of a boundary, the order of these operations does not matter. However, in the presence of a solid boundary, this symmetry breaks and collision and streaming operations, no longer commute with each other. Streaming is the generator of spatial translations while the boundary operator by construction must break translational invariance. Hence, in the presence of solid boundaries, the correct order of operations to calculate the flow forces, is not self evident. 
We describe two possible sequence of operations indicated by algorithm A and algorithm B in Fig. \ref{basicLadd}.
In Fig. \ref{basicResults1},\ref{basicResults2}, we show the effect of these two approaches on two test cases, that is, flow past a 2D circular cylinder and flow past a 3D NACA0012 airfoil, at moderate Reynolds numbers. These are standard non trivial test cases  extensively used for validation of numerical schemes and good amount of  benchmark experimental and computational data are available for the same \cite{succi1989, dennis, anderson,ilio}.  
To show that the discrepancy in drag values between the two algorithms, is generic, the force is computed with two discrete velocity models (D2Q9 and RD3Q41). 
The use of algorithm B is noticed in some literature\cite{ladd, caiazzo},
however, to the best of our knowledge, a detailed comparison between the two algorithms and an analysis as to why one is better than the other, is not presented in any work.  

\begin{figure}[htp]
	\includegraphics[scale=0.3]{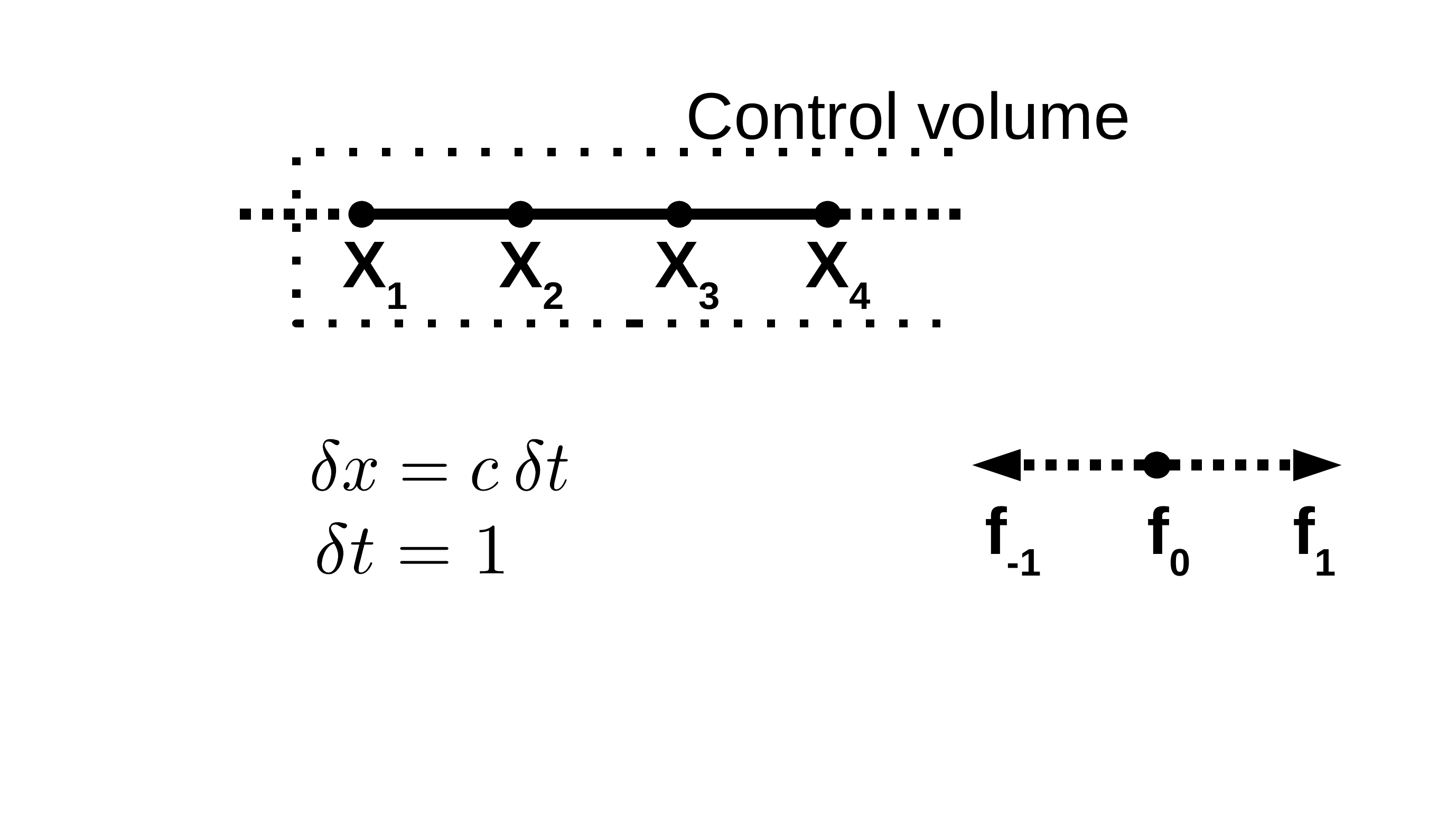}% Here is how to import EPS art
	\caption{\label{1d} A D1Q3 lattice with 4 grid points$(X_1,X_2,X_3,X_4)$, where $X_1$ is the boundary grid point of the control volume and $X_2$, $X_3$, $X_4$ are the bulk grid points. In the toy example, we assume that the grid is unbounded on the right end and continues to $X_5$, $X_6$ and so on.}
\end{figure}

In  this work, we investigate the difference in the two approaches, theoretically and numerically. In order to understand the differences, we formulate a discrete analogue of the Reynolds Transport Theorem (DRTT) in the discrete space-time setting of the lattice boltzmann equation and devise a simple and effective way for flow force calculation. The main idea in this approach is that  the complex-surfaced solid can be enclosed inside    a simpler cartesian-friendly bounding box, creating a control volume that facilitates balancing of momentum fluxes and accurate calculation of surface stresses on the solid boundary. The   method simplifies to MEA when the outer encapsulating surface falls directly onto the discrete solid surface \cite{oritz}. 
Hence, DRTT can be thought of as a modified MEA approach, built to tackle the errors that arise from the inability of cartesian grids to resolve complex surfaces, especially at low resolutions.  
To demonstrate this approach, we consider a 1D toy model as shown in Fig. \ref{1d}. The standard lattice Boltzmann equation (in one dimensional space) with a relaxation term is a two step evolution
\begin{align}
\begin{split}
& \tilde{f}_i(X,t)=f_i(X,t) + 2 \beta \left[f^{eq}_i(X,t) - f_i(X,t)\right],\\
& f_i(X + c_i \delta t, t+\delta t) = \tilde{f}_i(X,t).
\end{split}
\label{lbb1}
\end{align}
A more detailed description of notations used in lattice Boltzmann literature is given in the next section.
It would be convenient for further analysis to define the global momentum calculated over the control volume as,
\begin{equation}
J_X (t) = \sum_{q=1,2,3,4\cdots} \sum_{i} f_i (X_q, t) c_{i X}.  
\end{equation}
The evolution equation for global momentum derived from Eq.\eqref{lbb1}, for a semi-infinite control volume with only $X_1$ as the boundary point, simplifies to :
\begin{align}
\begin{split}
J_X(t+1) - J_X(t) = F_{1}(X_1,t+1)+ \left(f_{-1}-f_{1} +\tilde{f_{1}}\right)_{X_1,t},
\end{split}
\label{lbb2}
\end{align} 
where, $F_1(X_1,t+1)$ is the contribution from the boundary condition in a general case. In the simple case of a bounce-back boundary, if the boundary returns an un-collided distribution ($F_1(X_1,t+1) = f_{-1}(X_1,t)$), Eq.\eqref{lbb2} simplifies to,
\begin{equation}
J_X(t+1) - J_X(t) = \left( 2f_{-1} + 2 \beta \left[f_1^{eq} - f_1\right] \right)_{X_1,t}.
\label{lbb3}
\end{equation}

However global  momentum balance suggests that the RHS of Eq.\eqref{lbb3} is the total force that the boundary applies on the system. Using this example, it is easy to see that in Algorithm A, the force calculation routine (naive implementation of MEA for instance) misses out on the extra contribution, that is the second term on the RHS. If the procedure is followed for the collided populations, as per the recommendation of Algorithm B, one arrives at a simplified version of RTT,   
\begin{equation}
J_X(t+1) - J_X(t) = 2 \tilde{f}_{-1}(X_1,t) , 
\label{lbb4}
\end{equation}
where, $\tilde{f}_i$ represents a post-collision population.

To computationally validate our claims, we present simulations and convergence studies for NACA0012 at zero angle of attack for Reynolds numbers ranging from $10^2$ to $0.5 \times 10^6$.
The outline of this paper is as follows: In Sec. \ref{lbm}, we give a brief description of LBM. In Sec. \ref{force}, we review widely used methods for flow force calculation in LBM literature, namely stress integration and momentum exchange. In Sec. \ref{rtt}, we formulate a discrete analogue of the Reynolds tranport theorem to understand the reason behind the accuracy of Algorithm B. In Sec. \ref{results},  we demonstrate some numerical simulations to validate our findings.

\section{Lattice Boltzmann Model}
\label{lbm}
\begin{figure*}[htp]
	\centering
	\begin{tabular}[b]{c}
		\includegraphics[width=.4\linewidth]{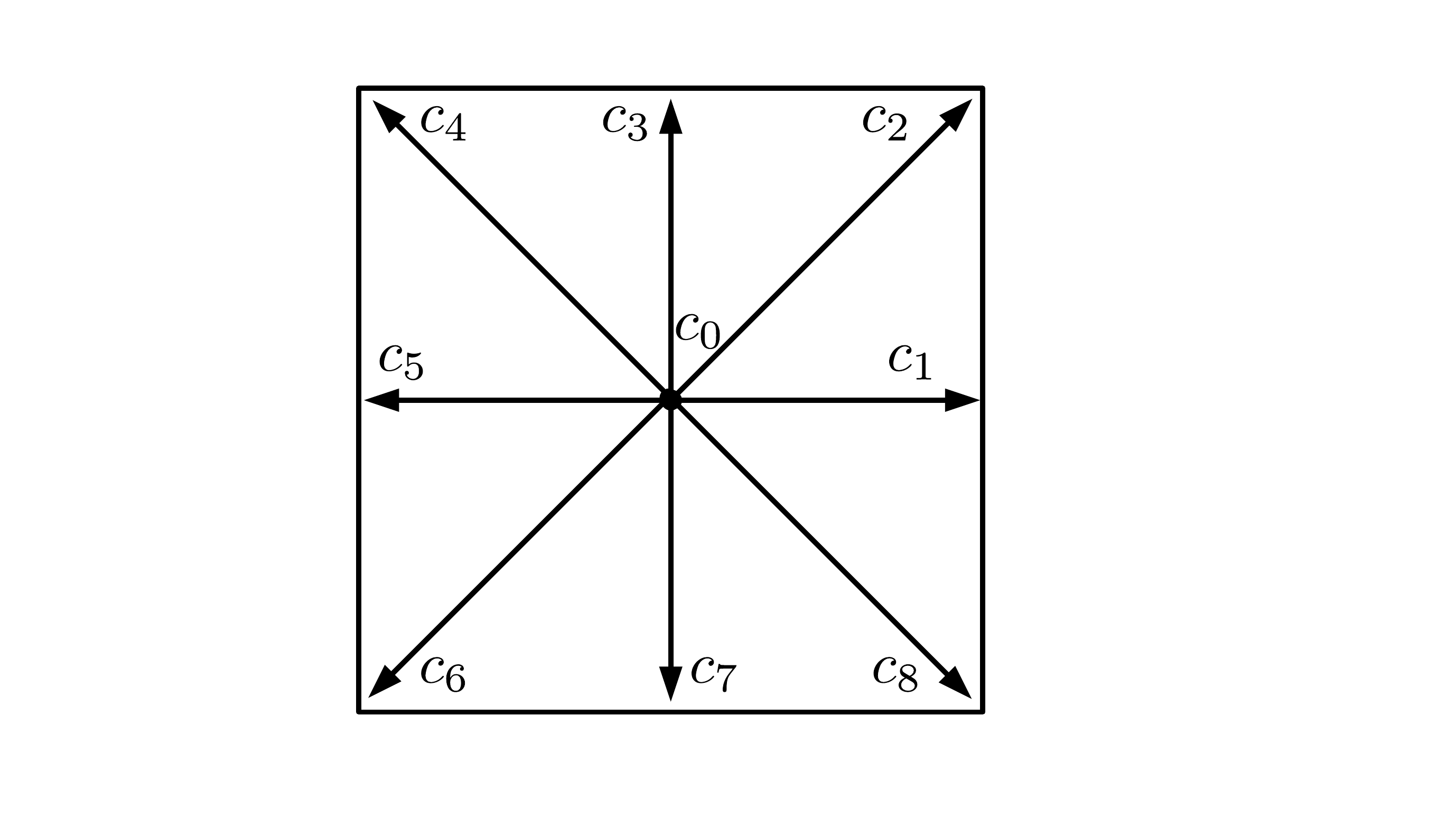} \\
		\small a)
	\end{tabular} \qquad
	\begin{tabular}[b]{c}
		\includegraphics[width=.5\linewidth]{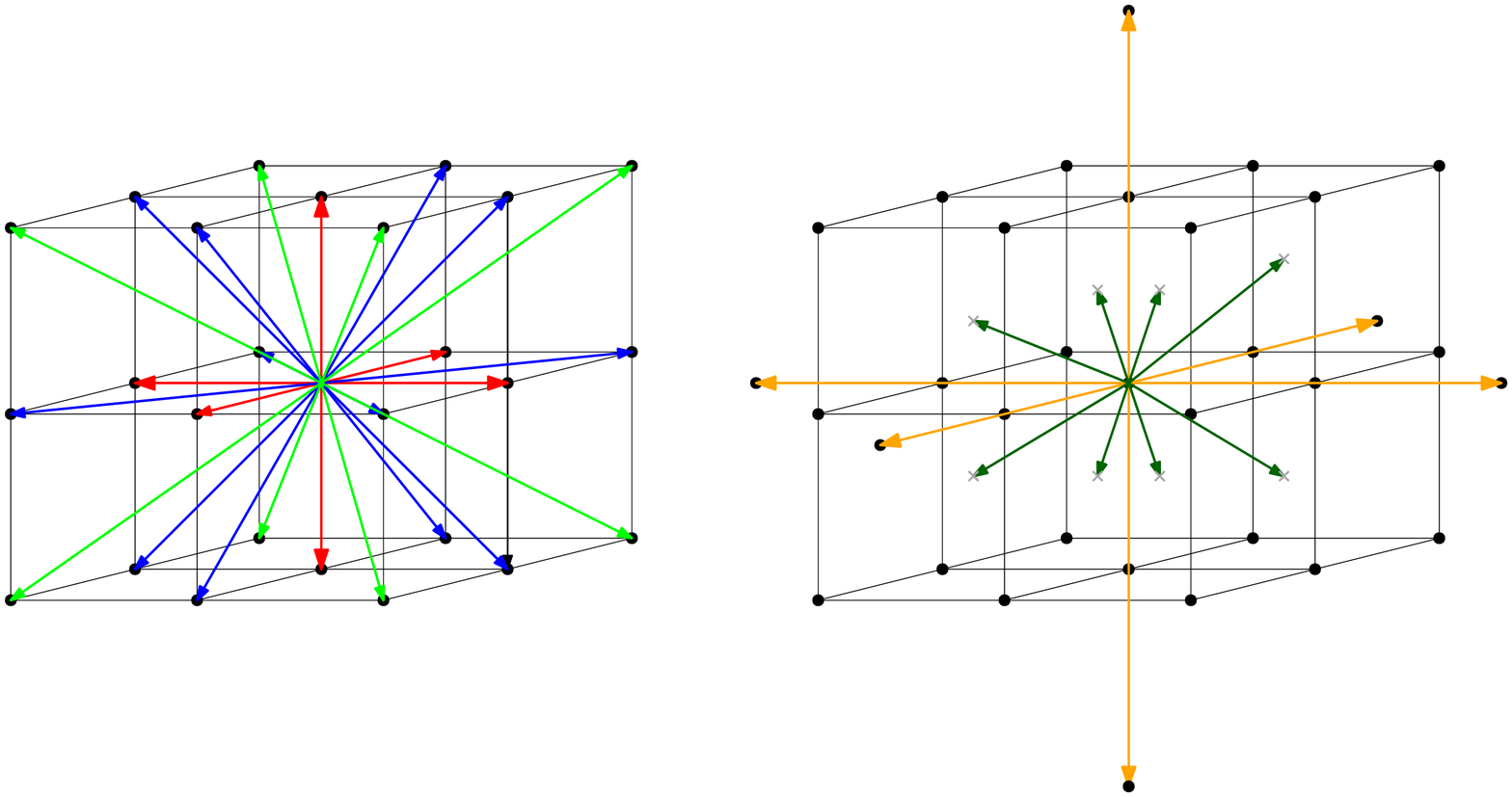} \\
		\small b)
	\end{tabular}
	\caption{\label{velvec}a)Energy shells of the D2Q9 model . b) Energy shells of the RD3Q41 model: fcc-1(blue), sc-1(red), bcc-1(light green) shown on a regular lattice along with sc-2(orange), bcc-1/2(dark green).}
\end{figure*} 

Lattice Boltzmann method represents hydrodynamics by a discrete space-time kinetic theory. The starting point for the method, is the construction of a discrete velocity set, ${\cal C}$,  consisting  of   $N_d$   discrete   velocities given by ${\mathbf c}_i$ ($i=1,2,\cdots N_d$).   
The set of basic variables, ${\mathbf f}$, for discrete kinetic theory are  populations $f_i({\mathbf x},t)$ of discrete velocities ${\mathbf c}_i$ ($i=1,2,\cdots N_d$), defined at location $\mathbf x$ and time $t$.  The hydrodynamic variables are the mass density $\rho$, the fluid velocity ${\mathbf u}$ and   scaled temperature $\theta$, 
defined in terms of Boltzmann constant $k_B$ and mass of the particle $m$, as $\theta = k_B T /m$. These macroscopic variables are related to the populations as, 
\begin{align}
\begin{split} 
\rho &= \sum_i^{N_d} f_i, \\
\rho u_{\alpha} &= \sum_i^{N_d} f_i  c_{i \alpha}, \\ 
 \rho u^2 + D \rho \theta &= \sum_i^{N_d} f_i c_i^2 ,
 \end{split} 
 \end{align}
 where, $D$ signifies the dimension of the setup.
 For these discrete velocity models, typically the kinetic equation is written with a single relaxation (BGK) term as,
\begin{equation}
\frac{\partial f_i}{\partial t} + \pmb c_i \cdot \pmb \nabla f_i = - \frac{1}{\tau} \left[ f_i - f^{eq}_i({\cal M}^{\rm Slow})\right],
\label{disc_boltzmann} 
\end{equation}
 where, $\tau$ is related to mean free time and $f^{eq}_i({\cal M}^{\rm Slow})$ is the discrete analog of Maxwell-Boltzmann distribution    with ${\cal M}^{\rm Slow}$ being, the set of slow hydrodynamic moments of populations. In continuous kinetic theory, the slow moments consists of mass, momentum and energy density (${\cal M}^{\rm Slow}=\left\{ \rho,  \rho {\mathbf{u}}, \rho u^2+3\rho \theta\right\}$), while in the discrete case one can often ignore energy conservation to focus on isothermal hydrodynamics only (${\cal M}^{\rm Slow}=\left\{ \rho, \rho {\mathbf{u}}\right\}$).
 The lattice boltzmann equation is obtained by further discretizing Eq.\eqref{disc_boltzmann} in space and time, as a sequence of discrete collisions followed by free flight(streaming)   as, 
\begin{align}
\begin{split}
&\tilde{f}_i(\pmb x,t) = f_i(\pmb x,t) + 2\beta \left[f^{eq}_i\left( \rho(\pmb x,t), \pmb u(\pmb x,t) \right) - f_i(\pmb x,t) \right],
\\
& f_i(\pmb x + \pmb c_i \delta t, t+\delta t) = \tilde{f}_i(\pmb x,t).
\end{split}
\label{CollideAdv} 
\end{align}
where $\beta = \delta t/(2\tau +\delta t)$. At this point, it needs to be pointed out that the same discrete evolution can also be written as free flight followed by collision as
\begin{align}
\begin{split}
&\hat{f}_i(\pmb x, t) =  {f}_i(\pmb x-   \pmb c_i \delta t,t),
\\
&f_i(\pmb x ,t+\delta t) = \hat{f}_i(\pmb x,t) + 2\beta \left[\hat{f}^{eq}_i\left( \rho(\pmb x,t), \pmb u(\pmb x,t) \right) - \hat{f}_i(\pmb x,t) \right].
  \end{split}
\label{AdvCollide} 
\end{align}
 It is straight-forward to see that the two versions are identical in absence of any solid boundaries. 
The main theoretical ingredient ensuring accurate hydrodynamics is the construction of the discrete equilibrium. Starting with a zero velocity equilibrium ($w_i>0$), a second order series approximation to the equilibrium is often written as,
 \begin{equation}
  f_i^{\rm eq }  = w_i\, \rho \left[1+  \frac{\left({\pmb u}\cdot  {\pmb c}_{i}  \right)}{\theta_0}+ \frac{ 1}{2 \theta_0^2} \left( \left({\pmb u}\cdot  {\pmb c}_{i}  \right)^2-  \theta_0   { u}^2    \right)\right]. 
 \end{equation} 
  Here, it is important that the zero velocity equilibrium at a fixed temperature is positive ($w_i>0$) and satisfies the condition that lower order moments are same as Maxwell-Boltzmann. In particular,
\begin{align}
\begin{split} 
    \sum_i^{N_d} w_i &=1, \\ 
    \sum_i^{N_d} w_i c_{i\alpha } c_{i \beta} &= \theta_0 \delta_{\alpha \beta}, \\
    \sum_i^{N_d} w_i c_{i\alpha } c_{i \beta} c_{i \kappa} c_{i \gamma} &=\theta_0^2 \Delta_{\alpha \beta \gamma \kappa}. 
\end{split}
\end{align}
These conditions on zero velocity equilibrium ensures, 
\begin{equation}
P_{\alpha \beta}^{\rm eq} \equiv \sum_i f_i^{\rm eq} c_{i \alpha} c_{i\beta}= \rho u_{\alpha}u_{ \beta} +\rho \theta_0\delta_{\alpha \beta}.
\end{equation}
Later with higher order lattices, the method was extended for finite, but subsonic Mach number case, by ensuring that  the contracted sixth order moment \cite{kolluru,atif} is,
\begin{equation}
\sum_i w_i c_i^2 c_{i\alpha } c_{i \beta} c_{i \kappa} c_{i \gamma} = 7 \theta_0^2 \Delta_{\alpha \beta \gamma \kappa}. 
\end{equation}
This allows one to write the equilibrium for a small departure from reference temperature, at least at zero velocity (second order in $\Delta \theta$) as,
\begin{align} 
\begin{split}
f_i^{(0)}     & =    w_i\Biggl[    1+  \frac{\Delta \theta}{2 }\left(   c_i -3 \right)+\frac{\Delta \theta^2 } {8 }
\left( c_i^2 -10 c_i +15    \right)\Biggr] .
\end{split}
\end{align}
which gives a
second order (in velocity) approximation to the equilibrium, 
\begin{equation}
\frac{f_i^{\rm Eq }}{\rho\,f^{(0)}}  =    1+  \frac{ {\pmb u}\cdot  {\pmb c}_{i}  }{\theta}+ \frac{  \left({\pmb u}\cdot  {\pmb c}_{i}  \right)^2-  \theta   {\pmb u}^2         -  \frac{ 5 \bar{R}  u^2}{ 6   +  1 5 \bar{R}} (c_{i}^2- 3 \theta) }{2 \theta^2},  
\end{equation}
The definition of $\bar{R}$ can be found in the cited reference \cite{kolluru}.   
The weights and discrete velocity set for two used models (D2Q9 and RD3Q41) are provided in Table \ref{tabD2Q9} and Table \ref{tabRD3Q41}. The energy shells for both the models is  
depicted in Fig. \ref{velvec}.

Arguably, the most crucial element of the lattice Boltzmann method is the boundary condition at a solid-fluid interface. Even though, the macroscopic boundary conditions are quite straight-forward, the mesoscopic boundary conditions on the discrete population($f_i(\pmb x,t)$), require some explanation. Historically, the two commonly used boundary conditions are bounce-back(BB) and diffusive boundary conditions \cite{ladd,aidun1998,noble, ansumali2007, shi2011}. In recent years, hybrid formulations like the diffused bounce-back boundary condition \cite{krithivasan, liu2021}, aimed at expanding the scope of application to a variety of problems, are also being used. In order to keep the message of this paper clear, we stick to the universally used BB boundary condition proposed by Ladd \cite{ladd}.   
The basic implementation of BB boundary condition, assumes that a molecule hits the wall and reverses its direction.
 
\begin{center}
\begin{table}[h]
\begin{tabular}{cccc} 
\hline \hline 
Shells    &  Discrete Velocities($c_i$)  &  Weight($w_i$) \\ \hline
zero & $\left(0,0\right)$ &$16/36$  \\ \hline
SC-$1$ & $ \left(\pm 1, 0  \right), \left(0, \pm 1 \right)$    & $4/36$   \\
\hline
FCC-$1$ & $  \left(\pm 1, \pm 1 \right)$ & $1/36$ \\
\hline
\end{tabular}
\caption{\label{tabD2Q9} Energy shells and their corresponding velocities with weights for D2Q9 with $\theta_0 = 1/3$.
}
\end{table}
\end{center}

\begin{table*}
	\begin{center}
		\begin{tabular}{cccc}
			\hline \hline
			Shells 
			& Discrete velocities($c_i$)   
			&  Weight($w_i$)                                                               \\ \hline
			zero
			& $\left( 0,0,0\right)$        
			&  $ \left(52 - 323 \theta_0 + 921 \theta_0^2 - 1036 \theta_0^3\right)/52$     \\
			SC-1
			&$\left(\pm 1, 0, 0  \right),\left(0, \pm 1, 0  \right),\left( 0, 0, \pm 1 \right) $ 
			&  $ \theta_0 \left(12 - 38 \theta_0 + 63 \theta_0^2\right)/39 $               \\
			SC-2
			&$\left(\pm 2, 0, 0  \right),\left(0, \pm 2, 0  \right),\left( 0, 0, \pm 2 \right) $ 
			&  $ \theta_0 \left(3 - 29 \theta_0 + 84 \theta_0^2\right)/312  $              \\
			FCC-1
			&$\left(\pm 1, \pm 1, 0  \right),\left(\pm 1, 0, \pm 1  \right),  \left( 0,\pm 1, \pm 1  \right)$ 
			& $  \theta_0 \left( 45 \theta_0 - 6 - 77 \theta_0^2\right)/26           $     \\
			BCC-1
			&$\left(\pm 1, \pm 1,  \pm 1  \right)$ 
			&   $ \theta_0 \left(20 - 163 \theta_0 + 378 \theta_0^2 \right)/312       $    \\
			BCC-0.5
			&$\left(\pm 0.5, \pm 0.5,  \pm 0.5  \right)$ 
			&  $ 8\theta_0 \left(4 - 17 \theta_0 + 21 \theta_0^2 \right)/39 $              \\ \hline
			\hline
		\end{tabular}
		\caption{Velocities and their corresponding weights for the RD3Q41 model
			with $\theta_0 = 0.2948964908710633$}
		\label{tabRD3Q41}
	\end{center}
\end{table*}

If $\pmb x\in \pmb x_f$ and $\pmb x + \pmb c_i \delta t \in \pmb x_s$ (where $\pmb x_f$ and $\pmb x_s$ denote the fluid and solid domain respectively), the boundary point, $\pmb x_b$, is assumed to lie at the midpoint of the vector joining $\pmb x_f$ and $\pmb x_s$, regardless of the physical position of the boundary (see Fig. \ref{basicLadd}). The distribution for the fluid points near the boundary is given by the boundary condition\cite{ladd},
\begin{align}
\begin{split} 
\text{Stationary wall : } & \tilde{f}_{\tilde{i}}(\pmb x_f,t+\delta t) = \tilde{f}_i(\pmb x_f, t)
\\
\text{Moving wall : } & \tilde{f}_{\tilde{i}}(\pmb x_f,t+\delta t) = \tilde{f}_i(\pmb x_f, t) - 2 w_i \rho \frac{\pmb c_i \cdot \pmb u_w}{c_s^2}
\end{split} 
\end{align} 
where $\tilde{i}$ is the direction opposite to $i$ and $\pmb u_w$ is the velocity of the solid wall. Here, $\tilde{f}$ represents the post-collision population. There are several higher-order interpolation based curved boundary conditions in literature that address the drawbacks in accuracy of the midway bounce-back boundary
condition when dealing with curved surfaces\cite{mei2000}. However, the essential ideas of this paper remain unchanged even with higher-order interpolation based schemes and can be transferred to the same.

 \section{Force Computation in LBM}
\label{force}
 Very often fluid simulations require 
 accurate determination of forces experienced by curved objects. 
 In this section, we begin by briefly reviewing the two widely used force evaluation schemes  in the \textit{•}context of LBM.
For fluid simulations with  any method, an intuitive way of calculating forces  is  to perform integration of the total stress on the contact surface. This method and all its variants are referred to as stress integration (SI) algorithms.
In terms of the unit normal $\pmb n$ pointing out of the solid boundary $\partial \Omega$,  the stress tensor on the boundary of a solid is  
\begin{equation}
\pmb F = \int_{\partial \Omega} dA \> \pmb n \cdot (-p \pmb I + \rho \nu [(\pmb \nabla : \pmb u)+(\pmb \nabla : \pmb u)^T])
\label{stress_int}
\end{equation}
where   the pressure term $p=  \rho \theta$ ($\theta=\theta_0$ for isothermal models)  can be computed easily at every grid point.  However, the deviatoric part ($\sigma_{\alpha \beta}$) of the stress tensor involves  the velocity gradient tensor $\partial_\alpha u_\beta$, that needs to be approximated via finite differences which typically  introduces additional error.   This  inaccuracy in calculation of velocity gradient tensor is circumvented in LBM by  observation that the   deviatoric stresses is evaluated as the second moment of   the non-equilibrium part of the  population \cite{filippova1998}
\begin{equation}
\sigma_{\alpha \beta} = \left(1 - \beta \right)  \Sigma_i [f_i(x,t) - f_i^{eq}(x,t)] \left(c_{i \alpha} c_{i \beta} - \frac{\theta_0}{D} \delta_{\alpha \beta}\right). 
\end{equation} 
 
However, LBM being a cartesian grid based method, another uncertainty in force estimation   is added by the calculation of a surface normal ($\pmb n$)  for complex geometries. Thus, for a  typical flow setup the SI calculation results have lower accuracy in contrast to the alternate  momentum exchange method \cite{mei2002}. 

A  computationally efficient and simple method is the momentum exchange (ME) algorithm \cite{ladd} where, the total forces  are computed as pairwise  sum of   momentum difference  
in discrete directions  as populations bounces back from body surface near
boundary points. In its basic version
 (see Fig \ref{basicLadd}), boundary is  approximated quite effectively in a staircase manner  at the middle of every
fluid-solid links, each of which crosses the boundary and connects a fluid node.    
The total force experienced by the solid object is given by,
\begin{equation}
\pmb F_b = \sum_b - \left[f_{\tilde{i}}(\pmb x_f,t)\pmb c_{\tilde{i}} - f_i(\pmb x_f,t)\pmb c_i\right],
\end{equation}
where, $\tilde{i}$ is the direction opposite to $i$ and the summation runs over all fluid points next to discrete boundary. 
The ambiguity pointed out in the introduction of the paper, regarding the choice of algorithm (see Fig.\ref{basicLadd}), plays an important role here. 
The momentum exchange algorithm is a limit to the more general Reynolds Transport Theorem \cite{oritz}. In the next section we derive the discrete version of the Reynolds Transport Theorem and arrive at simplistic force evaluation routine for complex shaped objects.  
 
\section{Discrete Reynolds Transport Theorem}
\label{rtt}
\begin{figure*}
	\includegraphics[scale=0.45]{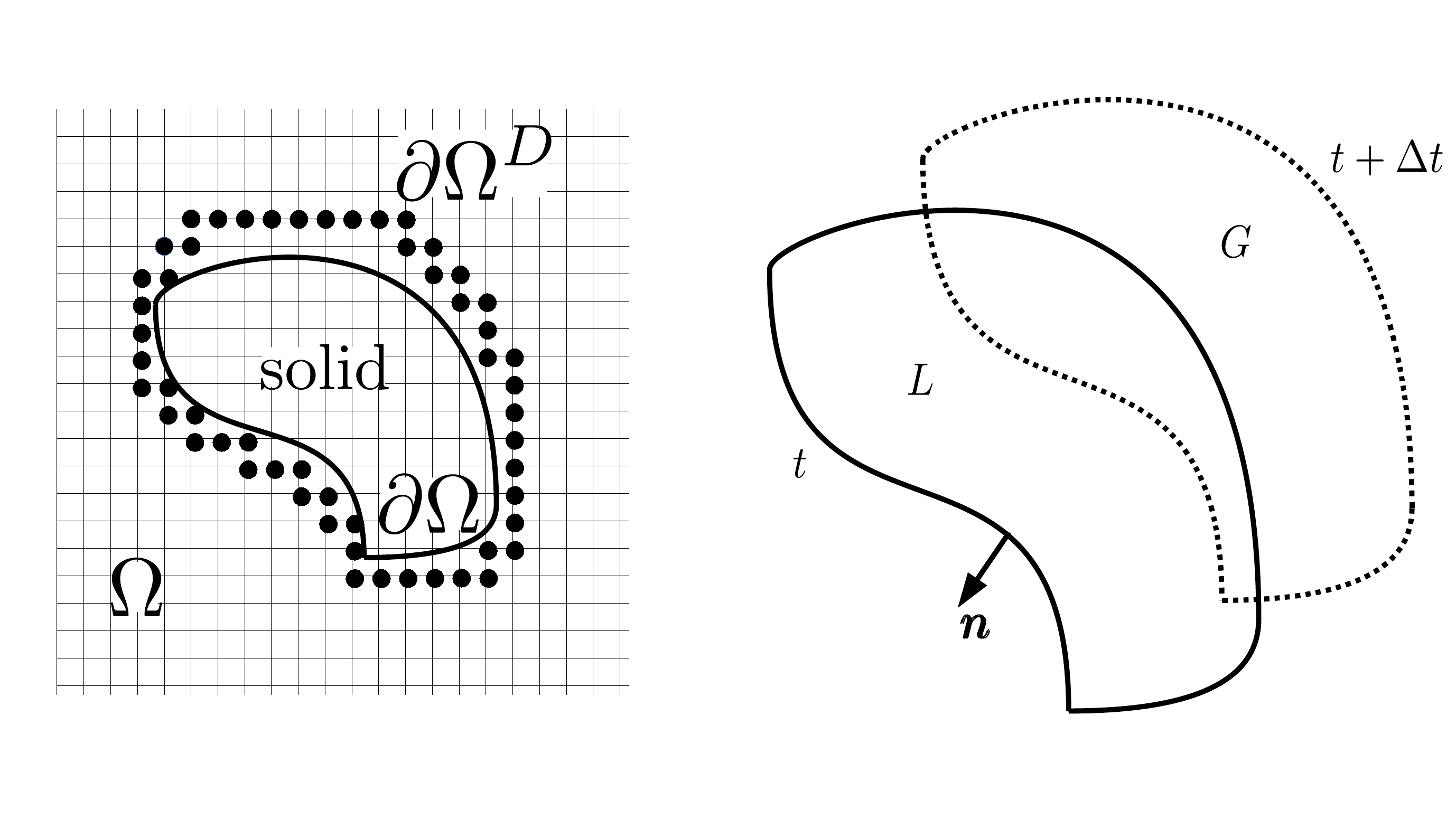}% Here is how to import EPS art
	\caption{\label{drtt1} Left : A solid object submerged in a fluid marked by $\Omega$ on a Cartesian grid. The circular dots represent the fluid nodes that make up the discrete boundary ($\partial \Omega^D$) analogous to the continuous solid boundary ($\partial \Omega$).
		\\ 
		Right : system evolving from $t$ to $t+\Delta t$.}
\end{figure*}

Consider a complex shaped solid object with a continuous boundary 
marked by $\partial \Omega$ as in Fig. \ref{drtt1}, placed on a Cartesian grid. 
The solid is surrounded by fluid nodes ($\Omega$) and the discrete analogue the boundary is represented by $\partial \Omega^D$. 
The total momentum  for all the fluid grid nodes in the bulk and boundary, is given by,
\begin{align}
\begin{split}
J^\Omega_\alpha (t) &= \sum_{\pmb x \in \Omega} \sum_{i} c_{i \alpha} f_i(\pmb x, t),
\label{rtt1}
\end{split}  
\end{align} 

The  momentum balance for the fluid points as the system evolves from $t$ to $t+\Delta t$, implies
\begin{equation}
J^\Omega_\alpha (t+\Delta t) = J^\Omega_\alpha (t) + J^G_\alpha(t+\Delta t) - J^L_\alpha(t)
\label{rtt2}
\end{equation}
where, $J^G_\alpha(t+\Delta t)$ and $J^L_\alpha(t)$ represents the gained and lost momentum at the boundary \cite{oritz}, as seen in Fig.\ref{drtt1}. It must be noted that $J^G_\alpha$ is calculated over the boundary grid nodes where population is added to the system and $J^L_\alpha$ is calculated over the boundary grid nodes where population leaves the system, when the system displaces from $t$ to $t+\Delta t$. The gained momentum is simply written as,
\begin{align}
\begin{split}
J^G_\alpha(t+\Delta t) &= \sum_{\pmb x\in \partial \Omega^D} \left[ \sum_{i \ni c_in_i > 0} c_{i \alpha} f_i(\pmb x,t+\Delta t) \right],
\\
&= \sum_{\pmb x\in \partial \Omega^D} \left[ \sum_{i \ni c_in_i > 0} c_{i \alpha} \tilde{f}_i(\pmb x - \pmb c \Delta t,t) \right],
\label{rtt3}
\end{split}  
\end{align}
where, $\tilde{f}_i$ denotes a collided population. As per the discussion in the introduction section of this article, there are two ways of calculating $J^L_\alpha(t)$, based on the choice of algorithm (A or B), 
\begin{align}
\begin{split}
\text{Algorithm A :   }& J^L_\alpha (t) = \sum_{\pmb x \in \partial \Omega^D} \left[ \sum_{i \ni c_in_i \leq 0} c_{i \alpha} f_i(\pmb x,t) \right],    
\\
\text{Algorithm B :   }& J^L_\alpha (t) = \sum_{\pmb x \in \partial \Omega^D} \left[ \sum_{i \ni c_in_i \leq 0} c_{i \alpha} \tilde{f}_i(\pmb x,t) \right].
\label{rtt4}
\end{split}
\end{align}
The choice of Algorithm B simplifies Eq.\eqref{rtt2} to give us the complete momentum balance for fluid nodes as,
\begin{align}
\begin{split}
J^\Omega_\alpha (t+\Delta t) - J^\Omega_\alpha (t) =
& \sum_{\pmb x\in \partial \Omega^D} \left[ \sum_{i \ni c_in_i > 0} c_{i \alpha} \tilde{f}_i(\pmb x - \pmb c \Delta t,t) \right] 
\\
-& \sum_{\pmb x \in \partial \Omega^D} \left[ \sum_{i \ni c_in_i \leq 0} c_{i \alpha} \tilde{f}_i(\pmb x,t) \right] 
\label{rtt5}
\end{split}  
\end{align} 
The boundary surface contribution on RHS of the Eq.\eqref{rtt5}, is exactly the momentum exchange algorithm proposed by Ladd \cite{ladd}, whereas the choice of Algorithm A would lead to a slightly more complicated expression. Using Newton's second law, one can estimate the forces exerted by the fluid, on submerged solid objects as,
\begin{align}
\begin{split}
F_\alpha (t) = &-\sum_{\pmb x\in \partial \Omega^D} \left[\sum_{i \ni c_in_i > 0} c_{i \alpha} \tilde{f}_i(\pmb x - \pmb c \Delta t,t) \right] 
\\
&+ \sum_{\pmb x\in \partial \Omega^D} \left[\sum_{i \ni c_in_i \leq 0} c_{i \alpha} \tilde{f}_i(\pmb x,t)   \right].
\end{split}
\end{align} 
\begin{figure*}
	\includegraphics[scale=0.45]{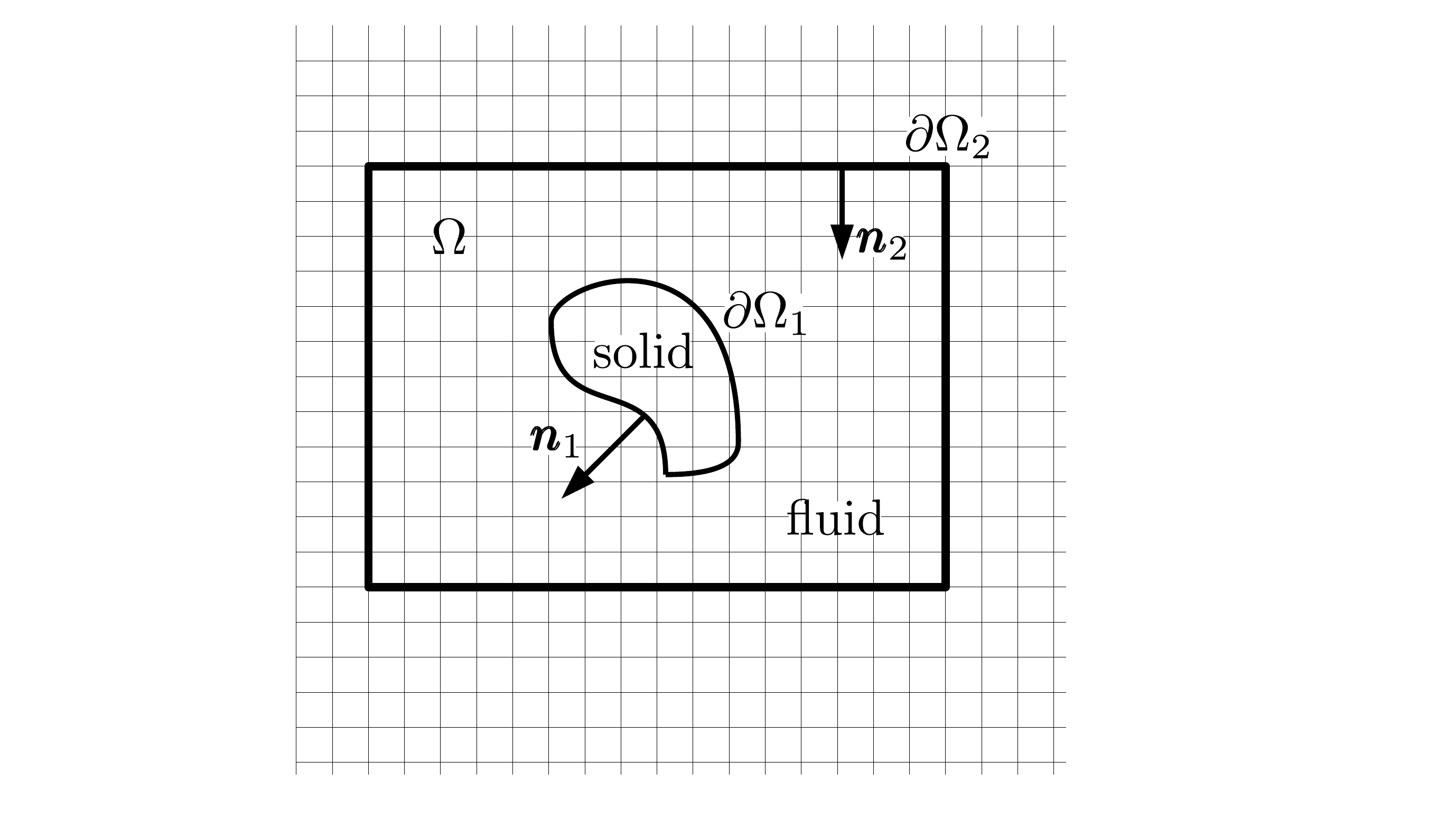}% Here is how to import EPS art
	\caption{\label{drtt2} A complex shaped solid object placed inside a control volume. The solid surface is marked by $\partial \Omega_1$ and the outer rectangular control volume surface is marked by $\partial \Omega_2$  }
\end{figure*}
The first term in the force formula is dictated by the boundary condition used in the algorithm. For a simple bounce-back boundary condition, the force formula simplifies to,
\begin{equation}
F_\alpha (t) = 2 \sum_{\pmb x\in \partial \Omega^D} \left[ \sum_{i \ni c_in_i \leq 0} c_{i \alpha} \tilde{f}_i(\pmb x,t)   \right].
\end{equation}

We extend the derivation carried out for a single boundary to a system with two boundaries as given in Fig.\ref{drtt2}, where the fluid nodes are given by $\Omega$, and are sandwiched between the two boundaries $\partial \Omega_1$ and $\partial \Omega_2$, such that the momentum balance gives,
\begin{align}
\begin{split}
J^{\Omega}_\alpha (t+\Delta t) - J^{\Omega}_\alpha (t) =
&\sum_{\pmb x\in \partial \Omega_2} \left[ \sum_{i \ni c_in2_i > 0} c_{i \alpha} \tilde{f}_i(\pmb x - \pmb c \Delta t,t) \right] 
\\
- &\sum_{\pmb x\in \partial \Omega_2} \left[ \sum_{i \ni c_in2_i \leq 0} c_{i \alpha} \tilde{f}_i(\pmb x,t)   \right]
\\
+ & \sum_{\pmb x\in \partial \Omega_1} \left[ \sum_{i \ni c_in1_i > 0} c_{i \alpha} \tilde{f}_i(\pmb x - \pmb c \Delta t,t) \right] 
\\
- & \sum_{\pmb x\in \partial \Omega_1} \left[ \sum_{i \ni c_in1_i \leq 0} c_{i \alpha} \tilde{f}_i(\pmb x,t)   \right]
\label{rtt7}
\end{split}
\end{align}
The construction of the outer boundary $\partial \Omega_2$ is done in such a way that $\pmb n_2$ and location of $\partial \Omega_2$ is trivial to resolve on a lattice. A rectangular bounding box aligned with the grid is a good example of the same. 
This construction helps us bypass calculations at the solid surface ($\partial \Omega_1$) by rearranging the Eq.\eqref{rtt7} to,
\begin{align}
\begin{split}
F_{\alpha}(t) = \>\> & J^{\Omega}_\alpha (t) -  J^{\Omega}_\alpha (t+\Delta t)   
\\
 + &\sum_{\pmb x\in \partial \Omega_2} \left[ \sum_{i \ni c_in2_i > 0} c_{i \alpha} \tilde{f}_i(\pmb x - \pmb c \Delta t,t) \right]
 \\ 
 - & \sum_{\pmb x\in \partial \Omega_2} \left[ \sum_{i \ni c_in2_i \leq 0} c_{i \alpha} \tilde{f}_i(\pmb x,t)   \right],
\label{rtt7}
\end{split}
\end{align}
where, $F_\alpha(t)$ is the force exerted by the fluid on the surface $\partial \Omega_1$. We label this force evaluation method as DRTT and conduct simulations in the next section to validate the method.
\section{Results}
\label{results}
%\subsection{Flow past 2D circular cylinder}
As a first example, we consider the flow past a circular cylinder in two dimensions, which is a regularly used benchmark case in computational fluid dynamics. 
The most basic calculation of the pressure coefficient on the surface of the cylinder involves calculation over boundary fluid nodes without any kind of extrapolation. The pressure coefficient at one boundary fluid node is given by,
\begin{equation}
C_p = \frac{p-p_{\infty}}{\frac{1}{2}\rho U^2} ,
\end{equation}
where, $p_{\infty}$ is the far upstream pressure.
\begin{figure*}[htp]
	\centering
	\begin{tabular}[b]{c}
		\includegraphics[width=.35\linewidth]{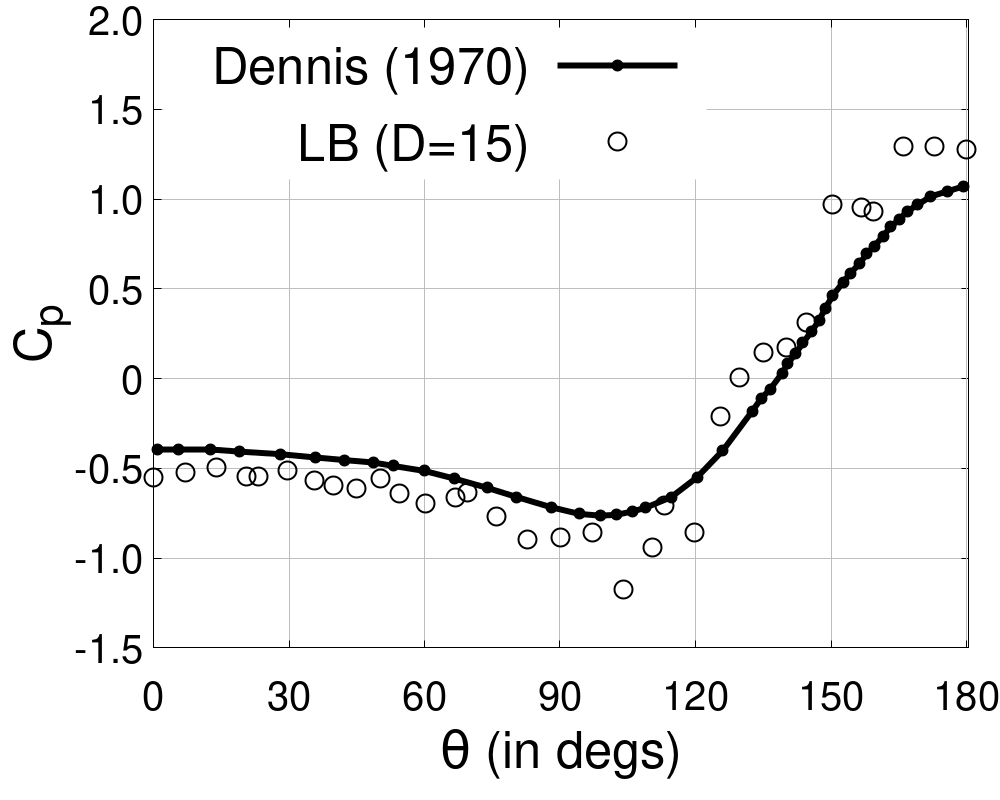} \\
		\small Case a)
	\end{tabular} \qquad
	\begin{tabular}[b]{c}
		\includegraphics[width=.39\linewidth]{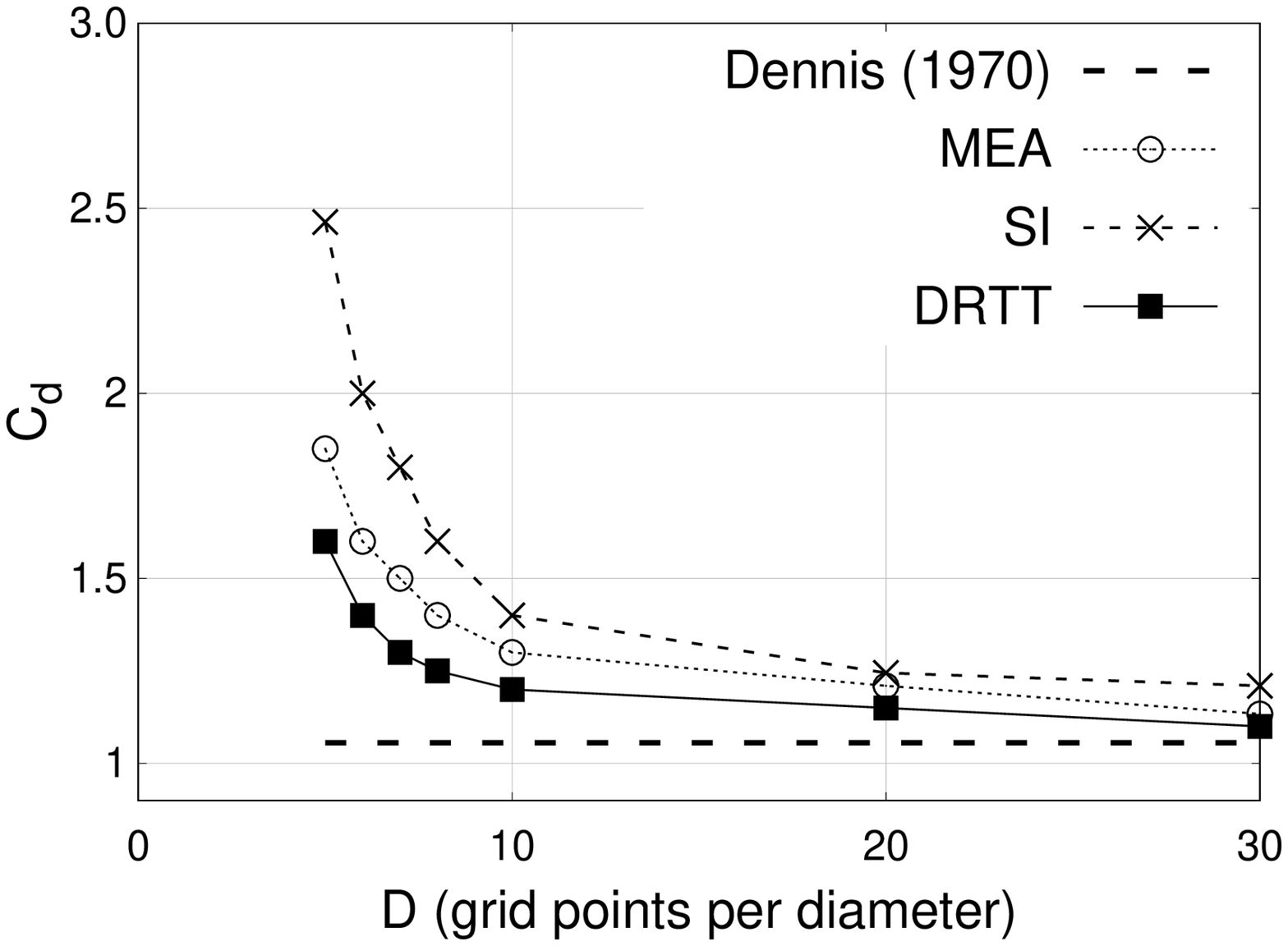} \\
		\small Case b)
	\end{tabular}
	\caption{\label{cp_cylinder} The flow past a circular cylinder in 2 dimension is a good bench-marking test case for lattice boltzmann methods \cite{succi1989}. Fig. \ref{cp_cylinder}a) shows the coefficient of pressure ($C_p$) vs $\theta$ for Re$=100$ case. Fig. \ref{cp_cylinder}b) compares the drag coefficient ($C_d$) as a function of number of grid point per diameter for momentum exchange(ME), stress integration(SI) and discrete RTT methods (DRTT).}
\end{figure*} 

Fig. \ref{cp_cylinder} shows the coefficient of pressure vs $\theta$ for Re$=100$ case. We see that as expected, the even at a low resolution of $D=15$, the $C_p$ curve is a close match with literature. Once the correctness of $C_p$ is established, the flow force is calculated using all the three methods discussed in the previous section. 
The drag coefficient over a circular cylinder of diameter D is given by,
\begin{equation}
C_d = \frac{2|F_x|}{\rho U^2 D} ,
\end{equation}
where, $x$ is taken to be the direction along the flow. The force term ($F_x$) is calculated using popular methods (MEA and SI), along with with the discrete Reynolds transport theorem (DRTT) algorithm.
We do the convergence test for all the methods for Re$=100$ and the results are given in Fig. \ref{cp_cylinder}. The aspect ratio for the computational domain is taken to be $12:1$ with the length and width of the domain as $60D$ and $5D$. The cylinder of diameter $D$ is placed at $15D$ from the inlet. We can clearly see that SI does the worst among all the methods at a given resolution and henceforth we shall not discuss it in the following sections on flow past a NACA airfoil.

\begin{figure}[htp]
	\includegraphics[scale=0.45]{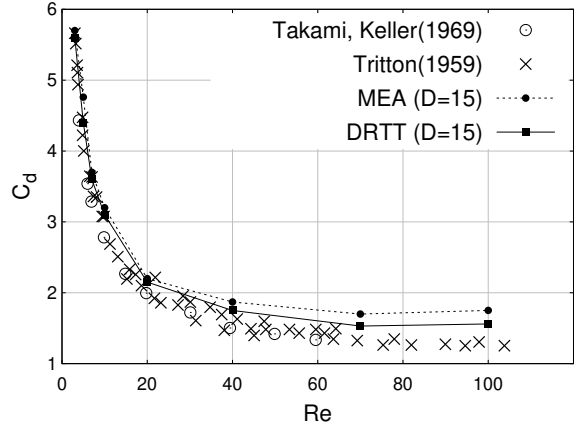}% Here is how to import EPS art
	\caption{\label{cd_re_cylinder} The computational domain size is kept fixed with $D=15$, while the Drag coefficient ($C_d$) is measured for several values of Reynolds numbers. The results of MEA and DRTT are then compared with literature. }
\end{figure}

The simulations are run for a range of parameters, in order to study the dependence of the drag coefficient on the Reynolds number. As seen in Fig. \ref{cd_re_cylinder}, DRTT does much better than MEA at a given resolution. One can also notice that the difference between DRTT and MEA becomes more prominent at high Re where the flow starts to become unsteady.

\begin{figure}[htp]
	\includegraphics[scale=0.45]{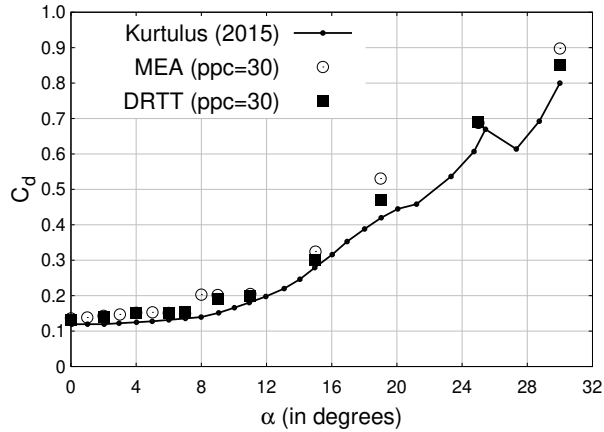}% Here is how to import EPS art
	\caption{\label{naca_Cd_alpha} Flow past a NACA0012 airfoil in two dimensions, using a D2Q9 lattice model and $30$ points per chord length. The coefficient of drag ($C_d$) plotted against angle of attack ($\alpha$) at Re=1000.}
\end{figure}

\subsection{Flow past 2D NACA0012 airfoil \\(Re=1000 benchmark)}
\begin{figure*}[htp]
	\centering
	\begin{tabular}[b]{c}
		\includegraphics[width=.28\linewidth]{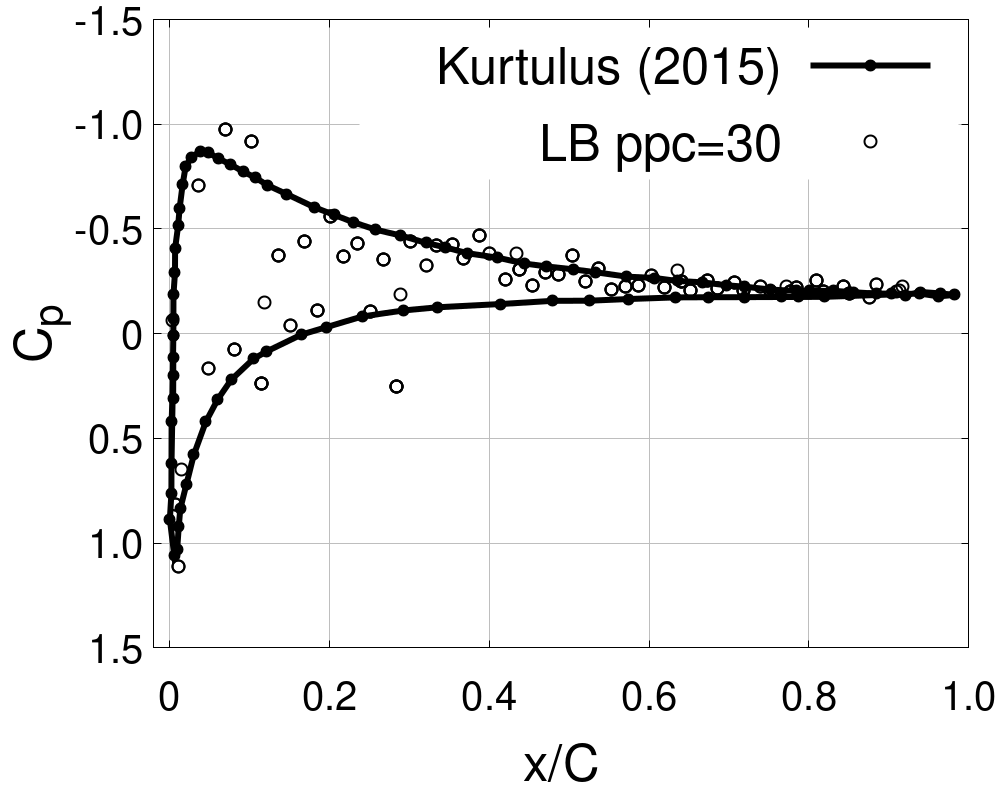} \\
		\small a)AoA = 7degs
	\end{tabular} \qquad
	\begin{tabular}[b]{c}
		\includegraphics[width=.28\linewidth]{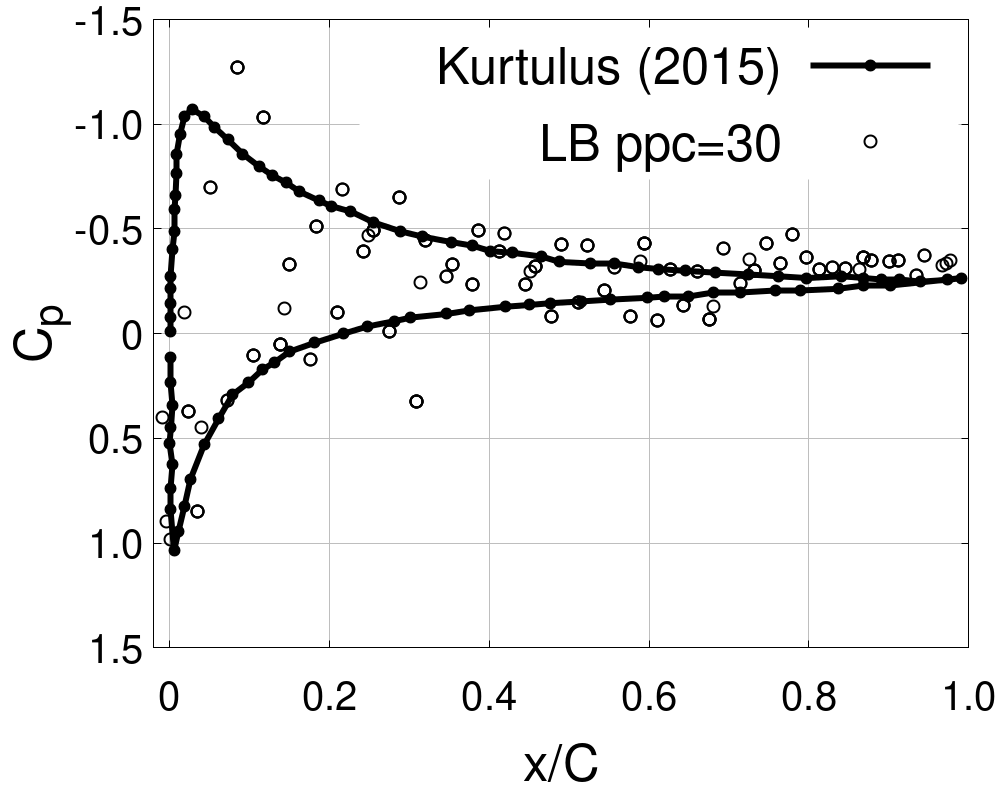} \\
		\small b)AoA = 9degs
	\end{tabular} \qquad
	\begin{tabular}[b]{c}
		\includegraphics[width=.28\linewidth]{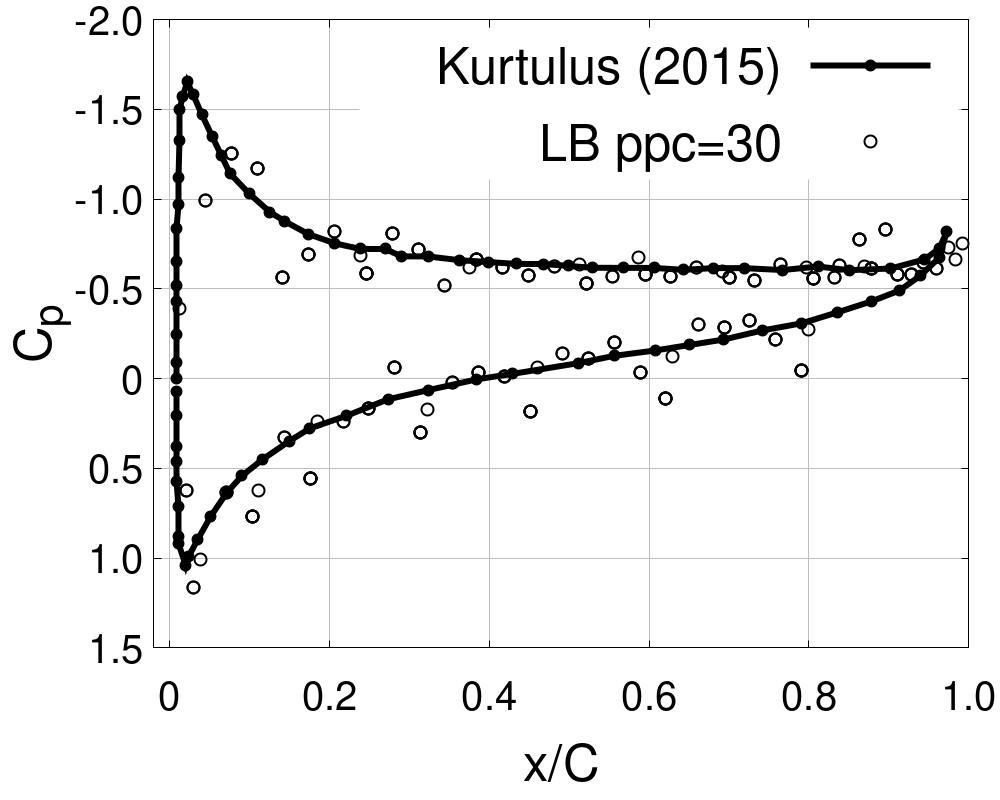} \\
		\small c)AoA = 15degs
	\end{tabular}
	\caption{\label{naca1000_cp} Flow past a NACA0012 airfoil in two dimensions for Reynolds=1000 \cite{ilio}. The coefficient of pressure ($C_p$) is calculated over the boundary fluid nodes for a computational domain size of 30 points per chord length. This is repeated for three distinct angles of attack ($\alpha = 7,9,15)^0$ and the results are matched with literature.}
\end{figure*}
The airfoil of interest in our work is the NACA0012 airfoil, that has been extensively used for validation cases for turbulence models.
NACA0012 airfoil is a skewed geometry with different grid requirements in x and y directions. This makes it an important case study for algorithms like lattice Boltzmann that use uniform meshes ($\Delta x = \Delta y$).
In order to validate our code, we reviewed the coefficient of pressure ($C_p$) measured over the airfoil surface at a Reynolds number of $1000$ and three angles of attack ($\alpha$) \cite{ilio}. Fig. \ref{naca1000_cp} shows us the match with literature. We plot for angle of attack$=(7,9,15)^0$  respectively and with a resolution of 30 points per chord.

\begin{figure}[htp]
	\includegraphics[scale=0.19]{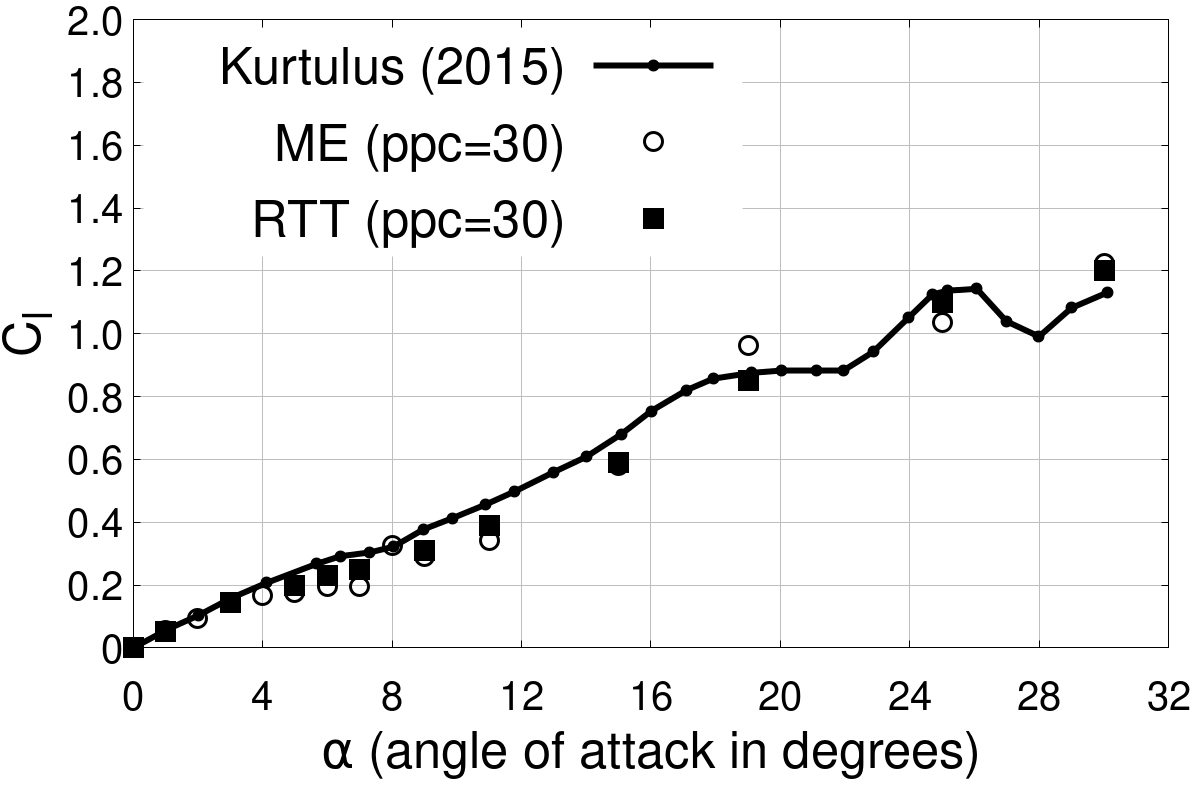}% Here is how to import EPS art
	\caption{\label{naca_Cl_alpha} Flow past a NACA0012 airfoil in two dimensions for Reynolds=1000. The coefficient of lift ($C_l$) plotted against angle of attack ($\alpha$).}
\end{figure}

The computational domain size is measured in the chord length ($C$) of the airfoil, such that, the number of grid points per chord length is given by $ppc$. The aspect ratio of the rectangular box is taken to be $12:1$. The length of the domain (in the direction of the flow) is taken to be $60 \times C$ with the airfoil placed symmetrically at $15 \times C$ from the inlet.
For the next set of results, simulations are run for a Reynolds of 1000 but the angle of attack is varied starting from 0 and finishing at 30 degrees. The drag and lift coefficients are calculated using both the MEA and DRTT flow force calculation methods. Both these methods are carried out for a resolution of 30 points per chord and the results are given in Fig. \ref{naca_Cd_alpha} and Fig. \ref{naca_Cl_alpha} respectively.

The lift coefficient ($C_l$) is given by,
\begin{equation}
C_l = \frac{2|F_y|}{\rho U^2 D}.
\end{equation}
We can see that MEA and DRTT do almost equally well at low angles of attack. But as the angle of attack rises and the flow begins to separate, DRTT is able to more accurately calculate the drag and lift values. We attribute this to the fact that DRTT, being a volumetric method, does not have to resolve the complex boundary of the airfoil at low grid resolution.

\begin{figure}[htp]
	\includegraphics[scale=0.45]{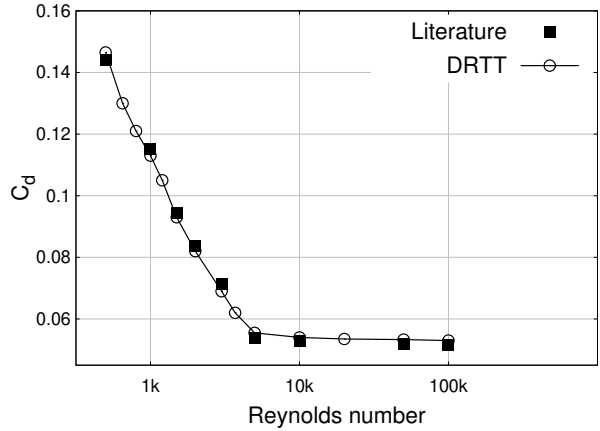}% Here is how to import EPS art
	\caption{\label{naca_cd_Re} The coefficient of drag from a 3 dimensional simulation of NACA0012 using RD3Q41 lattice boltzmann model. Converged results have been plotted }
\end{figure}

\subsection{Flow past 3D NACA0012 airfoil}
\begin{figure}[htp]
	\includegraphics[scale=0.45]{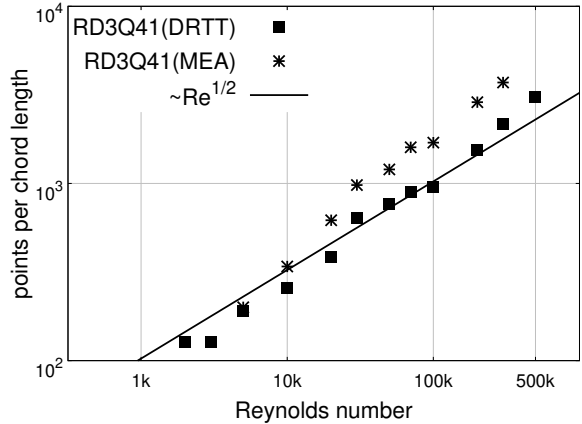}% Here is how to import EPS art
	\caption{\label{naca_3} Total number of computational points required to correctly measure the drag values for NACA0012 using the DRTT approach. The x-axis is the Reynolds number and the y-axis is the points per chord length of NACA0012.  }
\end{figure}

The final set of simulations for this study are for a 3 dimensional case of the NACA0012 airfoil. We conducted this at a fixed angle of attack of zero degrees but increased the Reynolds from a small value of $100$ to $0.5$ million. A 41 velocity lattice boltzmann model \cite{kolluru} was used for the 3 dimensional simulations. The goal of the study was to validate the DRTT approach for high Re values and also compare its behaviour with MEA.
The computational domain size is measured in the chord length of the airfoil ($C$). The number of grid points taken for $1C$ is given by $ppc$ (points per chord length).The aspect ratio is fixed to 10:1 in the x and y directions. The number of points in the z direction was varied with Reynolds number in order to accommodate a developing flow in the z direction at very high Reynolds. Table \ref{comp_details} gives details of the computational grid requirements for a converged result. The resolution demands (measured in points per chord) for a converged result vary as a function of Reynolds number. We assign an error percentage of 2 percent in order to call a result converged and tabulate the results in Table \ref{comp_details}. The converged $C_d$ vs $Re$ curve as seen in Fig. \ref{naca_cd_Re} shows a good match with literature. 
Fig. \ref{naca_3} shows a close match with a $Re^{1/2}$ scaling which can be explained using boundary layer theory.

\section{Conclusions}
In this work we revisited the force evaluation methodologies in the lattice Boltzmann method. 
The first half of the work emphasized on the non commutativity of the streaming and 
collision operations when dealing with flows around solid objects. 
This breakdown of commutativity opens up two possibilities for force 
evaluation, as shown in Fig. \ref{basicLadd}, the details of which have not been fully clarified in literature. Subsequent theoretical and computational analysis showed the superiority of Algorithm B over Algorithm A. In the process of establishing the aforementioned claim, the authors have suggested a simple and elegant force evaluation routine called DRTT. The elegance of DRTT lies in its compatibility with cartesian grid based methods like lattice Boltzmann, where the accurate resolution of complex shaped geometries is a major issue.
The DRTT routine is compared with the extensively used momentum-exchange method 
for a variety of flow problems, including flow past a two-dimensional cylinder and airfoil (NACA0012) 
and flow past a three-dimensional NACA0012 airfoil.
\section{Acknowledgments}
The authors are thankful to PK Kolluru and C Thantanapally, from SankhyaSutra Labs for their valuable inputs.
SS wishes to acknowledge funding from the European Research
Council under the Horizon 2020 Programme Grant Agreement n. 739964 ``COPMAT").

\begin{center}
\begin{table}[h]
\begin{tabular}{cccccc}
\\
\hline \hline 
Re                & Points per   &   & Points in   &  $\%$error in $C_d$   &  $\%$error in $C_d$   \\ 
                  & chord length &   & z-direction &  (MEA)                &  (DRTT) \\\hline
$1.0 \times 10^3$              &  128         &       & 12        &    1.75                   &   1.25       \\ \hline
$1.0 \times 10^4$            &  256         &       & 12        &  2.67                     &   1.74      \\ \hline
$0.1 \times 10^6$       &  960         &       & 24        &  3.1                     &   1.97       \\ \hline
$0.2 \times 10^6$       &  1536        &       & 48        &  3.22                     &   1.87        \\ \hline
$0.4 \times 10^6$       &  2816        &       & 48        &  5.16                     &   1.95        \\ \hline
$0.5 \times 10^6$       &  3072        &       & 60        &  9.21                     &   2.06         \\ \hline
\end{tabular}
\caption{\label{comp_details} The list of grid size requirements for a 3 dimensional NACA0012 simulation using a RD3Q41 model. The aspect ratio of the computational domain is taken to be $10:1$ with the length of the channel being $10\times$points per chord.}
\end{table}
\end{center}

\section{Appendix}
In this section we present a basic implementation of the discrete Reynolds transport theorem for flow past a two dimensional object. The size of the control volume ($\Omega$ as seen in Fig. \ref{drtt2}), for the simple case of 2 dimensional cylinder, is taken to be $1.5D \times 1.5D$ and the corners are marked as $cv_{x1}, cv_{x2}, cv_{y1}, cv_{y2}$. 

The basic algorithm for the force calculation using the DRTT approach is as follows:
\begin{equation*} 
\begin{split}
\xrightarrow{} & \text{Mark a rectangular DRTT control volume} (\Omega) \\
\xrightarrow{} & \text{Mark boundary fluid nodes around object} (\partial \Omega_1) \\
\xrightarrow{} & \text{Calculate total momentum in control volume} (\Omega) \\\\
& \hspace{2cm} for(i=cv_{x1}  , i \leq cv_{x2}, i++)\\
& \hspace{2cm} for(j=cv_{y1}+1, j<     cv_{y2}, j++) \\
& \hspace{2cm} J^{\Omega}_\alpha (t) += \sum_i c_{i \alpha} f_i(x,y,t)   
\\\\ 
\xrightarrow{} & \hspace{2cm} for(\text{entire fluid domain})\\
& \hspace{3cm} \text{Collide} 
\\\\
\xrightarrow{} & \text{Inlet-Outlet boundary conditions} \\
\xrightarrow{} & \text{Force calculation on the outer rectangular surface}\\
& \text{of the control volume} (\partial \Omega_2)\\
& \hspace{2cm} + \sum_{\pmb x\in \partial \Omega_2} \left[ \sum_{i \ni c_in2_i > 0} c_{i \alpha} \tilde{f}_i(\pmb x - \pmb c \Delta t,t) \right]
\\ 
& \hspace{2cm} - \sum_{\pmb x\in \partial \Omega_2} \left[ \sum_{i \ni c_in2_i \leq 0} c_{i \alpha} \tilde{f}_i(\pmb x,t)   \right]
\end{split}    
\end{equation*}
\begin{align} 
\begin{split}
\xrightarrow{} & \hspace{2cm} for(\text{entire fluid domain})\\
& \hspace{3cm} \text{Advection} 
\\\\
\xrightarrow{} & \text{Spectral boundary condition on the top-bottom walls} \\
& \text{and bounce-back on the object boundary}\\
\\
\xrightarrow{} & \text{Calculate total momentum in control volume} (\Omega) \\\\
& \hspace{2cm} for(i=cv_{x1}  , i \leq cv_{x2}, i++)\\
& \hspace{2cm} for(j=cv_{y1}+1, j<     cv_{y2}, j++) \\
& \hspace{2cm} J^{\Omega}_\alpha (t+\Delta t) += \sum_i c_{i \alpha} f_i(x,y,t+\Delta t)
\\
\xrightarrow{} & \text{DRTT force calculation as per Eq.} \eqref{rtt7} :\\
&\hspace{2cm} F_{\alpha}(t) =  J^{\Omega}_\alpha (t) -  J^{\Omega}_\alpha (t+\Delta t)   
\\
& \hspace{2cm} + \sum_{\pmb x\in \partial \Omega_2} \left[ \sum_{i \ni c_in2_i > 0} c_{i \alpha} \tilde{f}_i(\pmb x - \pmb c \Delta t,t) \right]
\\ 
& \hspace{2cm} - \sum_{\pmb x\in \partial \Omega_2} \left[ \sum_{i \ni c_in2_i \leq 0} c_{i \alpha} \tilde{f}_i(\pmb x,t)   \right]
\end{split}    
\end{align}

\vspace{100cm}
 
 \bibliography{references}% Produces the bibliography via BibTeX.

\begin{thebibliography}{33}%
\makeatletter
\providecommand \@ifxundefined [1]{%
 \@ifx{#1\undefined}
}%
\providecommand \@ifnum [1]{%
 \ifnum #1\expandafter \@firstoftwo
 \else \expandafter \@secondoftwo
 \fi
}%
\providecommand \@ifx [1]{%
 \ifx #1\expandafter \@firstoftwo
 \else \expandafter \@secondoftwo
 \fi
}%
\providecommand \natexlab [1]{#1}%
\providecommand \enquote  [1]{``#1''}%
\providecommand \bibnamefont  [1]{#1}%
\providecommand \bibfnamefont [1]{#1}%
\providecommand \citenamefont [1]{#1}%
\providecommand \href@noop [0]{\@secondoftwo}%
\providecommand \href [0]{\begingroup \@sanitize@url \@href}%
\providecommand \@href[1]{\@@startlink{#1}\@@href}%
\providecommand \@@href[1]{\endgroup#1\@@endlink}%
\providecommand \@sanitize@url [0]{\catcode `\\12\catcode `\$12\catcode
  `\&12\catcode `\#12\catcode `\^12\catcode `\_12\catcode `\%12\relax}%
\providecommand \@@startlink[1]{}%
\providecommand \@@endlink[0]{}%
\providecommand \url  [0]{\begingroup\@sanitize@url \@url }%
\providecommand \@url [1]{\endgroup\@href {#1}{\urlprefix }}%
\providecommand \urlprefix  [0]{URL }%
\providecommand \Eprint [0]{\href }%
\providecommand \doibase [0]{https://doi.org/}%
\providecommand \selectlanguage [0]{\@gobble}%
\providecommand \bibinfo  [0]{\@secondoftwo}%
\providecommand \bibfield  [0]{\@secondoftwo}%
\providecommand \translation [1]{[#1]}%
\providecommand \BibitemOpen [0]{}%
\providecommand \bibitemStop [0]{}%
\providecommand \bibitemNoStop [0]{.\EOS\space}%
\providecommand \EOS [0]{\spacefactor3000\relax}%
\providecommand \BibitemShut  [1]{\csname bibitem#1\endcsname}%
\let\auto@bib@innerbib\@empty
%</preamble>
\bibitem [{\citenamefont {Higuera}\ \emph {et~al.}(1989)\citenamefont
  {Higuera}, \citenamefont {Succi},\ and\ \citenamefont {Benzi}}]{higuera1989}%
  \BibitemOpen
  \bibfield  {author} {\bibinfo {author} {\bibfnamefont {F.}~\bibnamefont
  {Higuera}}, \bibinfo {author} {\bibfnamefont {S.}~\bibnamefont {Succi}},\
  and\ \bibinfo {author} {\bibfnamefont {R.}~\bibnamefont {Benzi}},\
  }\href@noop {} {\bibfield  {journal} {\bibinfo  {journal} {EPL (Europhysics
  Letters)}\ }\textbf {\bibinfo {volume} {9}},\ \bibinfo {pages} {345}
  (\bibinfo {year} {1989})}\BibitemShut {NoStop}%
\bibitem [{\citenamefont {Higuera}\ and\ \citenamefont
  {Succi}(1989)}]{higuera1989-2}%
  \BibitemOpen
  \bibfield  {author} {\bibinfo {author} {\bibfnamefont {F.}~\bibnamefont
  {Higuera}}\ and\ \bibinfo {author} {\bibfnamefont {S.}~\bibnamefont
  {Succi}},\ }\href@noop {} {\bibfield  {journal} {\bibinfo  {journal} {EPL
  (Europhysics Letters)}\ }\textbf {\bibinfo {volume} {8}},\ \bibinfo {pages}
  {517} (\bibinfo {year} {1989})}\BibitemShut {NoStop}%
\bibitem [{\citenamefont {Thantanapally}\ \emph {et~al.}(2013)\citenamefont
  {Thantanapally}, \citenamefont {Patil}, \citenamefont {Succi},\ and\
  \citenamefont {Ansumali}}]{chakri2013}%
  \BibitemOpen
  \bibfield  {author} {\bibinfo {author} {\bibfnamefont {C.}~\bibnamefont
  {Thantanapally}}, \bibinfo {author} {\bibfnamefont {D.~V.}\ \bibnamefont
  {Patil}}, \bibinfo {author} {\bibfnamefont {S.}~\bibnamefont {Succi}},\ and\
  \bibinfo {author} {\bibfnamefont {S.}~\bibnamefont {Ansumali}},\ }\href@noop
  {} {\bibfield  {journal} {\bibinfo  {journal} {Journal of Fluid Mechanics}\
  }\textbf {\bibinfo {volume} {728}} (\bibinfo {year} {2013})}\BibitemShut
  {NoStop}%
\bibitem [{\citenamefont {Singh}\ \emph {et~al.}(2011)\citenamefont {Singh},
  \citenamefont {Subramanian},\ and\ \citenamefont {Ansumali}}]{singh2013}%
  \BibitemOpen
  \bibfield  {author} {\bibinfo {author} {\bibfnamefont {S.}~\bibnamefont
  {Singh}}, \bibinfo {author} {\bibfnamefont {G.}~\bibnamefont {Subramanian}},\
  and\ \bibinfo {author} {\bibfnamefont {S.}~\bibnamefont {Ansumali}},\
  }\href@noop {} {\bibfield  {journal} {\bibinfo  {journal} {Philosophical
  Transactions of the Royal Society A: Mathematical, Physical and Engineering
  Sciences}\ }\textbf {\bibinfo {volume} {369}},\ \bibinfo {pages} {2301}
  (\bibinfo {year} {2011})}\BibitemShut {NoStop}%
\bibitem [{\citenamefont {Thampi}\ \emph {et~al.}(2013)\citenamefont {Thampi},
  \citenamefont {Ansumali}, \citenamefont {Adhikari},\ and\ \citenamefont
  {Succi}}]{thampi2013}%
  \BibitemOpen
  \bibfield  {author} {\bibinfo {author} {\bibfnamefont {S.~P.}\ \bibnamefont
  {Thampi}}, \bibinfo {author} {\bibfnamefont {S.}~\bibnamefont {Ansumali}},
  \bibinfo {author} {\bibfnamefont {R.}~\bibnamefont {Adhikari}},\ and\
  \bibinfo {author} {\bibfnamefont {S.}~\bibnamefont {Succi}},\ }\href@noop {}
  {\bibfield  {journal} {\bibinfo  {journal} {Journal of Computational
  Physics}\ }\textbf {\bibinfo {volume} {234}},\ \bibinfo {pages} {1} (\bibinfo
  {year} {2013})}\BibitemShut {NoStop}%
\bibitem [{\citenamefont {Ansumali}\ \emph {et~al.}(2007)\citenamefont
  {Ansumali}, \citenamefont {Karlin}, \citenamefont {Arcidiacono},
  \citenamefont {Abbas},\ and\ \citenamefont {Prasianakis}}]{ansumali2007}%
  \BibitemOpen
  \bibfield  {author} {\bibinfo {author} {\bibfnamefont {S.}~\bibnamefont
  {Ansumali}}, \bibinfo {author} {\bibfnamefont {I.}~\bibnamefont {Karlin}},
  \bibinfo {author} {\bibfnamefont {S.}~\bibnamefont {Arcidiacono}}, \bibinfo
  {author} {\bibfnamefont {A.}~\bibnamefont {Abbas}},\ and\ \bibinfo {author}
  {\bibfnamefont {N.}~\bibnamefont {Prasianakis}},\ }\href@noop {} {\bibfield
  {journal} {\bibinfo  {journal} {Physical review letters}\ }\textbf {\bibinfo
  {volume} {98}},\ \bibinfo {pages} {124502} (\bibinfo {year}
  {2007})}\BibitemShut {NoStop}%
\bibitem [{\citenamefont {Succi}(2001)}]{succi2001}%
  \BibitemOpen
  \bibfield  {author} {\bibinfo {author} {\bibfnamefont {S.}~\bibnamefont
  {Succi}},\ }\href@noop {} {\emph {\bibinfo {title} {The lattice Boltzmann
  equation: for fluid dynamics and beyond}}}\ (\bibinfo  {publisher} {Oxford
  university press},\ \bibinfo {year} {2001})\BibitemShut {NoStop}%
\bibitem [{\citenamefont {Succi}\ and\ \citenamefont
  {Succi}(2018)}]{succi2018lattice}%
  \BibitemOpen
  \bibfield  {author} {\bibinfo {author} {\bibfnamefont {S.}~\bibnamefont
  {Succi}}\ and\ \bibinfo {author} {\bibfnamefont {S.}~\bibnamefont {Succi}},\
  }\href@noop {} {\emph {\bibinfo {title} {The lattice Boltzmann equation: for
  complex states of flowing matter}}}\ (\bibinfo  {publisher} {Oxford
  University Press},\ \bibinfo {year} {2018})\BibitemShut {NoStop}%
\bibitem [{\citenamefont {Benzi}\ \emph {et~al.}(1992)\citenamefont {Benzi},
  \citenamefont {Succi},\ and\ \citenamefont {Vergassola}}]{benzi1992}%
  \BibitemOpen
  \bibfield  {author} {\bibinfo {author} {\bibfnamefont {R.}~\bibnamefont
  {Benzi}}, \bibinfo {author} {\bibfnamefont {S.}~\bibnamefont {Succi}},\ and\
  \bibinfo {author} {\bibfnamefont {M.}~\bibnamefont {Vergassola}},\
  }\href@noop {} {\bibfield  {journal} {\bibinfo  {journal} {Physics Reports}\
  }\textbf {\bibinfo {volume} {222}},\ \bibinfo {pages} {145} (\bibinfo {year}
  {1992})}\BibitemShut {NoStop}%
\bibitem [{\citenamefont {Aidun}\ and\ \citenamefont {Clausen}(2010)}]{aidun}%
  \BibitemOpen
  \bibfield  {author} {\bibinfo {author} {\bibfnamefont {C.~K.}\ \bibnamefont
  {Aidun}}\ and\ \bibinfo {author} {\bibfnamefont {J.~R.}\ \bibnamefont
  {Clausen}},\ }\href@noop {} {\bibfield  {journal} {\bibinfo  {journal}
  {Annual review of fluid mechanics}\ }\textbf {\bibinfo {volume} {42}},\
  \bibinfo {pages} {439} (\bibinfo {year} {2010})}\BibitemShut {NoStop}%
\bibitem [{\citenamefont {Chen}\ and\ \citenamefont {Doolen}(1998)}]{chen}%
  \BibitemOpen
  \bibfield  {author} {\bibinfo {author} {\bibfnamefont {S.}~\bibnamefont
  {Chen}}\ and\ \bibinfo {author} {\bibfnamefont {G.~D.}\ \bibnamefont
  {Doolen}},\ }\href@noop {} {\bibfield  {journal} {\bibinfo  {journal} {Annual
  review of fluid mechanics}\ }\textbf {\bibinfo {volume} {30}},\ \bibinfo
  {pages} {329} (\bibinfo {year} {1998})}\BibitemShut {NoStop}%
\bibitem [{\citenamefont {Ansumali}\ and\ \citenamefont
  {Karlin}(2002)}]{ansumali2002}%
  \BibitemOpen
  \bibfield  {author} {\bibinfo {author} {\bibfnamefont {S.}~\bibnamefont
  {Ansumali}}\ and\ \bibinfo {author} {\bibfnamefont {I.~V.}\ \bibnamefont
  {Karlin}},\ }\href@noop {} {\bibfield  {journal} {\bibinfo  {journal}
  {Physical Review E}\ }\textbf {\bibinfo {volume} {66}},\ \bibinfo {pages}
  {026311} (\bibinfo {year} {2002})}\BibitemShut {NoStop}%
\bibitem [{\citenamefont {Li}\ \emph {et~al.}(2004)\citenamefont {Li},
  \citenamefont {Lu}, \citenamefont {Fang},\ and\ \citenamefont {Qian}}]{li}%
  \BibitemOpen
  \bibfield  {author} {\bibinfo {author} {\bibfnamefont {H.}~\bibnamefont
  {Li}}, \bibinfo {author} {\bibfnamefont {X.}~\bibnamefont {Lu}}, \bibinfo
  {author} {\bibfnamefont {H.}~\bibnamefont {Fang}},\ and\ \bibinfo {author}
  {\bibfnamefont {Y.}~\bibnamefont {Qian}},\ }\href@noop {} {\bibfield
  {journal} {\bibinfo  {journal} {Physical Review E}\ }\textbf {\bibinfo
  {volume} {70}},\ \bibinfo {pages} {026701} (\bibinfo {year}
  {2004})}\BibitemShut {NoStop}%
\bibitem [{\citenamefont {Filippova}\ and\ \citenamefont
  {H{\"a}nel}(1998{\natexlab{a}})}]{olga}%
  \BibitemOpen
  \bibfield  {author} {\bibinfo {author} {\bibfnamefont {O.}~\bibnamefont
  {Filippova}}\ and\ \bibinfo {author} {\bibfnamefont {D.}~\bibnamefont
  {H{\"a}nel}},\ }\href@noop {} {\bibfield  {journal} {\bibinfo  {journal}
  {Journal of Computational physics}\ }\textbf {\bibinfo {volume} {147}},\
  \bibinfo {pages} {219} (\bibinfo {year} {1998}{\natexlab{a}})}\BibitemShut
  {NoStop}%
\bibitem [{\citenamefont {Inamuro}\ \emph {et~al.}(2000)\citenamefont
  {Inamuro}, \citenamefont {Maeba},\ and\ \citenamefont {Ogino}}]{inamuro}%
  \BibitemOpen
  \bibfield  {author} {\bibinfo {author} {\bibfnamefont {T.}~\bibnamefont
  {Inamuro}}, \bibinfo {author} {\bibfnamefont {K.}~\bibnamefont {Maeba}},\
  and\ \bibinfo {author} {\bibfnamefont {F.}~\bibnamefont {Ogino}},\
  }\href@noop {} {\bibfield  {journal} {\bibinfo  {journal} {International
  journal of multiphase flow}\ }\textbf {\bibinfo {volume} {26}},\ \bibinfo
  {pages} {1981} (\bibinfo {year} {2000})}\BibitemShut {NoStop}%
\bibitem [{\citenamefont {Ladd}(1994{\natexlab{a}})}]{ladd}%
  \BibitemOpen
  \bibfield  {author} {\bibinfo {author} {\bibfnamefont {A.~J.}\ \bibnamefont
  {Ladd}},\ }\href@noop {} {\bibfield  {journal} {\bibinfo  {journal} {Journal
  of fluid mechanics}\ }\textbf {\bibinfo {volume} {271}},\ \bibinfo {pages}
  {285} (\bibinfo {year} {1994}{\natexlab{a}})}\BibitemShut {NoStop}%
\bibitem [{\citenamefont {Mei}\ \emph {et~al.}(2002)\citenamefont {Mei},
  \citenamefont {Yu}, \citenamefont {Shyy},\ and\ \citenamefont
  {Luo}}]{mei2002}%
  \BibitemOpen
  \bibfield  {author} {\bibinfo {author} {\bibfnamefont {R.}~\bibnamefont
  {Mei}}, \bibinfo {author} {\bibfnamefont {D.}~\bibnamefont {Yu}}, \bibinfo
  {author} {\bibfnamefont {W.}~\bibnamefont {Shyy}},\ and\ \bibinfo {author}
  {\bibfnamefont {L.-S.}\ \bibnamefont {Luo}},\ }\href@noop {} {\bibfield
  {journal} {\bibinfo  {journal} {Physical Review E}\ }\textbf {\bibinfo
  {volume} {65}},\ \bibinfo {pages} {041203} (\bibinfo {year}
  {2002})}\BibitemShut {NoStop}%
\bibitem [{\citenamefont {Ladd}(1994{\natexlab{b}})}]{ladd1994}%
  \BibitemOpen
  \bibfield  {author} {\bibinfo {author} {\bibfnamefont {A.~J.}\ \bibnamefont
  {Ladd}},\ }\href@noop {} {\bibfield  {journal} {\bibinfo  {journal} {Journal
  of fluid mechanics}\ }\textbf {\bibinfo {volume} {271}},\ \bibinfo {pages}
  {311} (\bibinfo {year} {1994}{\natexlab{b}})}\BibitemShut {NoStop}%
\bibitem [{\citenamefont {Mei}\ \emph {et~al.}(2000)\citenamefont {Mei},
  \citenamefont {Shyy}, \citenamefont {Yu},\ and\ \citenamefont
  {Luo}}]{mei2000}%
  \BibitemOpen
  \bibfield  {author} {\bibinfo {author} {\bibfnamefont {R.}~\bibnamefont
  {Mei}}, \bibinfo {author} {\bibfnamefont {W.}~\bibnamefont {Shyy}}, \bibinfo
  {author} {\bibfnamefont {D.}~\bibnamefont {Yu}},\ and\ \bibinfo {author}
  {\bibfnamefont {L.-S.}\ \bibnamefont {Luo}},\ }\href@noop {} {\bibfield
  {journal} {\bibinfo  {journal} {Journal of Computational Physics}\ }\textbf
  {\bibinfo {volume} {161}},\ \bibinfo {pages} {680} (\bibinfo {year}
  {2000})}\BibitemShut {NoStop}%
\bibitem [{\citenamefont {Succi}\ \emph {et~al.}(1989)\citenamefont {Succi},
  \citenamefont {Foti},\ and\ \citenamefont {Higuera}}]{succi1989}%
  \BibitemOpen
  \bibfield  {author} {\bibinfo {author} {\bibfnamefont {S.}~\bibnamefont
  {Succi}}, \bibinfo {author} {\bibfnamefont {E.}~\bibnamefont {Foti}},\ and\
  \bibinfo {author} {\bibfnamefont {F.}~\bibnamefont {Higuera}},\ }\href@noop
  {} {\bibfield  {journal} {\bibinfo  {journal} {EPL (Europhysics Letters)}\
  }\textbf {\bibinfo {volume} {10}},\ \bibinfo {pages} {433} (\bibinfo {year}
  {1989})}\BibitemShut {NoStop}%
\bibitem [{\citenamefont {Dennis}\ and\ \citenamefont {Chang}(1970)}]{dennis}%
  \BibitemOpen
  \bibfield  {author} {\bibinfo {author} {\bibfnamefont {S.}~\bibnamefont
  {Dennis}}\ and\ \bibinfo {author} {\bibfnamefont {G.-Z.}\ \bibnamefont
  {Chang}},\ }\href@noop {} {\bibfield  {journal} {\bibinfo  {journal} {Journal
  of Fluid Mechanics}\ }\textbf {\bibinfo {volume} {42}},\ \bibinfo {pages}
  {471} (\bibinfo {year} {1970})}\BibitemShut {NoStop}%
\bibitem [{\citenamefont {Warren}\ \emph {et~al.}(1991)\citenamefont {Warren},
  \citenamefont {Anderson}, \citenamefont {Thomas},\ and\ \citenamefont
  {Krist}}]{anderson}%
  \BibitemOpen
  \bibfield  {author} {\bibinfo {author} {\bibfnamefont {G.}~\bibnamefont
  {Warren}}, \bibinfo {author} {\bibfnamefont {W.}~\bibnamefont {Anderson}},
  \bibinfo {author} {\bibfnamefont {J.}~\bibnamefont {Thomas}},\ and\ \bibinfo
  {author} {\bibfnamefont {S.}~\bibnamefont {Krist}},\ }in\ \href@noop {}
  {\emph {\bibinfo {booktitle} {10th Computational Fluid Dynamics
  Conference}}}\ (\bibinfo {year} {1991})\ p.\ \bibinfo {pages}
  {1592}\BibitemShut {NoStop}%
\bibitem [{\citenamefont {Di~Ilio}\ \emph {et~al.}(2018)\citenamefont
  {Di~Ilio}, \citenamefont {Chiappini}, \citenamefont {Ubertini}, \citenamefont
  {Bella},\ and\ \citenamefont {Succi}}]{ilio}%
  \BibitemOpen
  \bibfield  {author} {\bibinfo {author} {\bibfnamefont {G.}~\bibnamefont
  {Di~Ilio}}, \bibinfo {author} {\bibfnamefont {D.}~\bibnamefont {Chiappini}},
  \bibinfo {author} {\bibfnamefont {S.}~\bibnamefont {Ubertini}}, \bibinfo
  {author} {\bibfnamefont {G.}~\bibnamefont {Bella}},\ and\ \bibinfo {author}
  {\bibfnamefont {S.}~\bibnamefont {Succi}},\ }\href@noop {} {\bibfield
  {journal} {\bibinfo  {journal} {Computers \& Fluids}\ }\textbf {\bibinfo
  {volume} {166}},\ \bibinfo {pages} {200} (\bibinfo {year}
  {2018})}\BibitemShut {NoStop}%
\bibitem [{\citenamefont {Caiazzo}\ and\ \citenamefont {Junk}(2008)}]{caiazzo}%
  \BibitemOpen
  \bibfield  {author} {\bibinfo {author} {\bibfnamefont {A.}~\bibnamefont
  {Caiazzo}}\ and\ \bibinfo {author} {\bibfnamefont {M.}~\bibnamefont {Junk}},\
  }\href@noop {} {\bibfield  {journal} {\bibinfo  {journal} {Computers \&
  Mathematics with Applications}\ }\textbf {\bibinfo {volume} {55}},\ \bibinfo
  {pages} {1415} (\bibinfo {year} {2008})}\BibitemShut {NoStop}%
\bibitem [{\citenamefont {Giovacchini}\ and\ \citenamefont
  {Ortiz}(2015)}]{oritz}%
  \BibitemOpen
  \bibfield  {author} {\bibinfo {author} {\bibfnamefont {J.~P.}\ \bibnamefont
  {Giovacchini}}\ and\ \bibinfo {author} {\bibfnamefont {O.~E.}\ \bibnamefont
  {Ortiz}},\ }\href@noop {} {\bibfield  {journal} {\bibinfo  {journal}
  {Physical Review E}\ }\textbf {\bibinfo {volume} {92}},\ \bibinfo {pages}
  {063302} (\bibinfo {year} {2015})}\BibitemShut {NoStop}%
\bibitem [{\citenamefont {Kolluru}\ \emph {et~al.}(2020)\citenamefont
  {Kolluru}, \citenamefont {Atif}, \citenamefont {Namburi},\ and\ \citenamefont
  {Ansumali}}]{kolluru}%
  \BibitemOpen
  \bibfield  {author} {\bibinfo {author} {\bibfnamefont {P.~K.}\ \bibnamefont
  {Kolluru}}, \bibinfo {author} {\bibfnamefont {M.}~\bibnamefont {Atif}},
  \bibinfo {author} {\bibfnamefont {M.}~\bibnamefont {Namburi}},\ and\ \bibinfo
  {author} {\bibfnamefont {S.}~\bibnamefont {Ansumali}},\ }\href@noop {}
  {\bibfield  {journal} {\bibinfo  {journal} {Physical Review E}\ }\textbf
  {\bibinfo {volume} {101}},\ \bibinfo {pages} {013309} (\bibinfo {year}
  {2020})}\BibitemShut {NoStop}%
\bibitem [{\citenamefont {Atif}\ \emph {et~al.}(2018)\citenamefont {Atif},
  \citenamefont {Namburi},\ and\ \citenamefont {Ansumali}}]{atif}%
  \BibitemOpen
  \bibfield  {author} {\bibinfo {author} {\bibfnamefont {M.}~\bibnamefont
  {Atif}}, \bibinfo {author} {\bibfnamefont {M.}~\bibnamefont {Namburi}},\ and\
  \bibinfo {author} {\bibfnamefont {S.}~\bibnamefont {Ansumali}},\ }\href@noop
  {} {\bibfield  {journal} {\bibinfo  {journal} {Physical Review E}\ }\textbf
  {\bibinfo {volume} {98}},\ \bibinfo {pages} {053311} (\bibinfo {year}
  {2018})}\BibitemShut {NoStop}%
\bibitem [{\citenamefont {Aidun}\ \emph {et~al.}(1998)\citenamefont {Aidun},
  \citenamefont {Lu},\ and\ \citenamefont {Ding}}]{aidun1998}%
  \BibitemOpen
  \bibfield  {author} {\bibinfo {author} {\bibfnamefont {C.~K.}\ \bibnamefont
  {Aidun}}, \bibinfo {author} {\bibfnamefont {Y.}~\bibnamefont {Lu}},\ and\
  \bibinfo {author} {\bibfnamefont {E.-J.}\ \bibnamefont {Ding}},\ }\href@noop
  {} {\bibfield  {journal} {\bibinfo  {journal} {Journal of Fluid Mechanics}\
  }\textbf {\bibinfo {volume} {373}},\ \bibinfo {pages} {287} (\bibinfo {year}
  {1998})}\BibitemShut {NoStop}%
\bibitem [{\citenamefont {Noble}\ \emph {et~al.}(1995)\citenamefont {Noble},
  \citenamefont {Chen}, \citenamefont {Georgiadis},\ and\ \citenamefont
  {Buckius}}]{noble}%
  \BibitemOpen
  \bibfield  {author} {\bibinfo {author} {\bibfnamefont {D.~R.}\ \bibnamefont
  {Noble}}, \bibinfo {author} {\bibfnamefont {S.}~\bibnamefont {Chen}},
  \bibinfo {author} {\bibfnamefont {J.~G.}\ \bibnamefont {Georgiadis}},\ and\
  \bibinfo {author} {\bibfnamefont {R.~O.}\ \bibnamefont {Buckius}},\
  }\href@noop {} {\bibfield  {journal} {\bibinfo  {journal} {Physics of
  Fluids}\ }\textbf {\bibinfo {volume} {7}},\ \bibinfo {pages} {203} (\bibinfo
  {year} {1995})}\BibitemShut {NoStop}%
\bibitem [{\citenamefont {Shi}\ \emph {et~al.}(2011)\citenamefont {Shi},
  \citenamefont {Brookes}, \citenamefont {Yap},\ and\ \citenamefont
  {Sader}}]{shi2011}%
  \BibitemOpen
  \bibfield  {author} {\bibinfo {author} {\bibfnamefont {Y.}~\bibnamefont
  {Shi}}, \bibinfo {author} {\bibfnamefont {P.~L.}\ \bibnamefont {Brookes}},
  \bibinfo {author} {\bibfnamefont {Y.~W.}\ \bibnamefont {Yap}},\ and\ \bibinfo
  {author} {\bibfnamefont {J.~E.}\ \bibnamefont {Sader}},\ }\href@noop {}
  {\bibfield  {journal} {\bibinfo  {journal} {Physical Review E}\ }\textbf
  {\bibinfo {volume} {83}},\ \bibinfo {pages} {045701} (\bibinfo {year}
  {2011})}\BibitemShut {NoStop}%
\bibitem [{\citenamefont {Krithivasan}\ \emph {et~al.}(2014)\citenamefont
  {Krithivasan}, \citenamefont {Wahal},\ and\ \citenamefont
  {Ansumali}}]{krithivasan}%
  \BibitemOpen
  \bibfield  {author} {\bibinfo {author} {\bibfnamefont {S.}~\bibnamefont
  {Krithivasan}}, \bibinfo {author} {\bibfnamefont {S.}~\bibnamefont {Wahal}},\
  and\ \bibinfo {author} {\bibfnamefont {S.}~\bibnamefont {Ansumali}},\
  }\href@noop {} {\bibfield  {journal} {\bibinfo  {journal} {Physical Review
  E}\ }\textbf {\bibinfo {volume} {89}},\ \bibinfo {pages} {033313} (\bibinfo
  {year} {2014})}\BibitemShut {NoStop}%
\bibitem [{\citenamefont {Liu}\ and\ \citenamefont {Lee}(2021)}]{liu2021}%
  \BibitemOpen
  \bibfield  {author} {\bibinfo {author} {\bibfnamefont {G.}~\bibnamefont
  {Liu}}\ and\ \bibinfo {author} {\bibfnamefont {T.}~\bibnamefont {Lee}},\
  }\href@noop {} {\bibfield  {journal} {\bibinfo  {journal} {Computers \&
  Fluids}\ }\textbf {\bibinfo {volume} {220}},\ \bibinfo {pages} {104884}
  (\bibinfo {year} {2021})}\BibitemShut {NoStop}%
\bibitem [{\citenamefont {Filippova}\ and\ \citenamefont
  {H{\"a}nel}(1998{\natexlab{b}})}]{filippova1998}%
  \BibitemOpen
  \bibfield  {author} {\bibinfo {author} {\bibfnamefont {O.}~\bibnamefont
  {Filippova}}\ and\ \bibinfo {author} {\bibfnamefont {D.}~\bibnamefont
  {H{\"a}nel}},\ }\href@noop {} {\bibfield  {journal} {\bibinfo  {journal}
  {Journal of computational Physics}\ }\textbf {\bibinfo {volume} {147}},\
  \bibinfo {pages} {219} (\bibinfo {year} {1998}{\natexlab{b}})}\BibitemShut
  {NoStop}%
\end{thebibliography}%
\end{document}